\documentclass[sigplan,10pt]{acmart}
\usepackage{wrapfig}
\usepackage{comment}
\usepackage{colortbl}
\usepackage[normalem]{ulem}
\usepackage{hyperref}

\usepackage{amsthm} 

\usepackage{color,soul}
\usepackage{graphics}
\usepackage{hyperref}
\usepackage{lipsum}
\usepackage{tikz}
\usepackage{enumitem}	
\usepackage{listings} 
\usepackage{booktabs}	
\usepackage{lipsum}

\usepackage{capt-of}	
\usepackage{diagbox}
\usepackage{multirow}
\usepackage{todonotes}  
\usepackage{subcaption} 
\usepackage{caption}

\definecolor{refkey}{rgb}{1,0,0}
\definecolor{labelkey}{rgb}{0,1,0}

\definecolor{airforceblue}{rgb}{0.36, 0.54, 0.66}
\definecolor{applegreen}{rgb}{0.55, 0.71, 0.0}

\usepackage{mathrsfs}	
\usepackage{amsfonts}

\usepackage[export]{adjustbox} 

\usepackage[plain]{algorithm2e}
\usepackage{boldline}					

\usepackage{multirow}
\usepackage{cuted} 

\definecolor{frenzyorange}{RGB}{249, 158, 26}

\renewcommand{\paragraph}[1]{\vskip 3pt\noindent\textbf{#1 }}	 

  {\begin{list}{$\bullet$}%
     {\setlength{\parsep}{0pt}%
      \setlength{\topsep}{0pt}%
      \setlength{\itemsep}{2pt}}}%
  {\end{list}}
\newcommand\Noted[1]{} 

\definecolor{darkblue}{rgb}{0.0, 0.0, 0.55}
\definecolor{mygreen}{HTML}{ADFF2F}
\definecolor{mylightgray}{gray}{0.8}

\usepackage[inline]{trackchanges}
\addeditor{WC}

\newenvironment{myitemize}%
  {\begin{itemize}
	[leftmargin=0cm,
		itemindent=.3cm,
		labelwidth=\itemindent,
		labelsep=0pt,
		parsep=1pt,
		topsep=1pt,
		itemsep=1pt,
		align=left]
  }%
  {\end{itemize}}    

  {\begin{enumerate}
	[leftmargin=.cm,itemindent=.5cm,labelwidth=\itemindent,
		labelsep=0pt,
		parsep=1pt,
		topsep=1pt,
		itemsep=3pt,
		align=left]
  }%
  {\end{enumerate}}    

\newcommand\sect[1]{Section~\ref{sec:#1}}	

\newcommand{\sys}{Proto}

\makeatletter
\def\@copyrightspace{\relax}
\makeatother

\begin{document}

\acmYear{2025}\copyrightyear{2025}
\setcopyright{cc}
\setcctype[4.0]{by}
\acmConference[SOSP '25]{ACM SIGOPS 31st Symposium on Operating Systems Principles}{October 13--16, 2025}{Seoul, Republic of Korea}
\acmBooktitle{ACM SIGOPS 31st Symposium on Operating Systems Principles (SOSP '25), October 13--16, 2025, Seoul, Republic of Korea}
\acmDOI{10.1145/3731569.3764811}
\acmISBN{979-8-4007-1870-0/25/10}

\title{Proto: A Guided Journey through Modern OS Construction}


\author{Wonkyo Choe*, Rongxiang Wang*, Afsara Benazir*, Felix Xiaozhu Lin}
\affiliation{%
  \institution{University of Virginia}
  \country{USA}
}



\date{}


\begin{strip}
	\centering
    \vspace{-20pt}
	\includegraphics[width=0.98\textwidth]{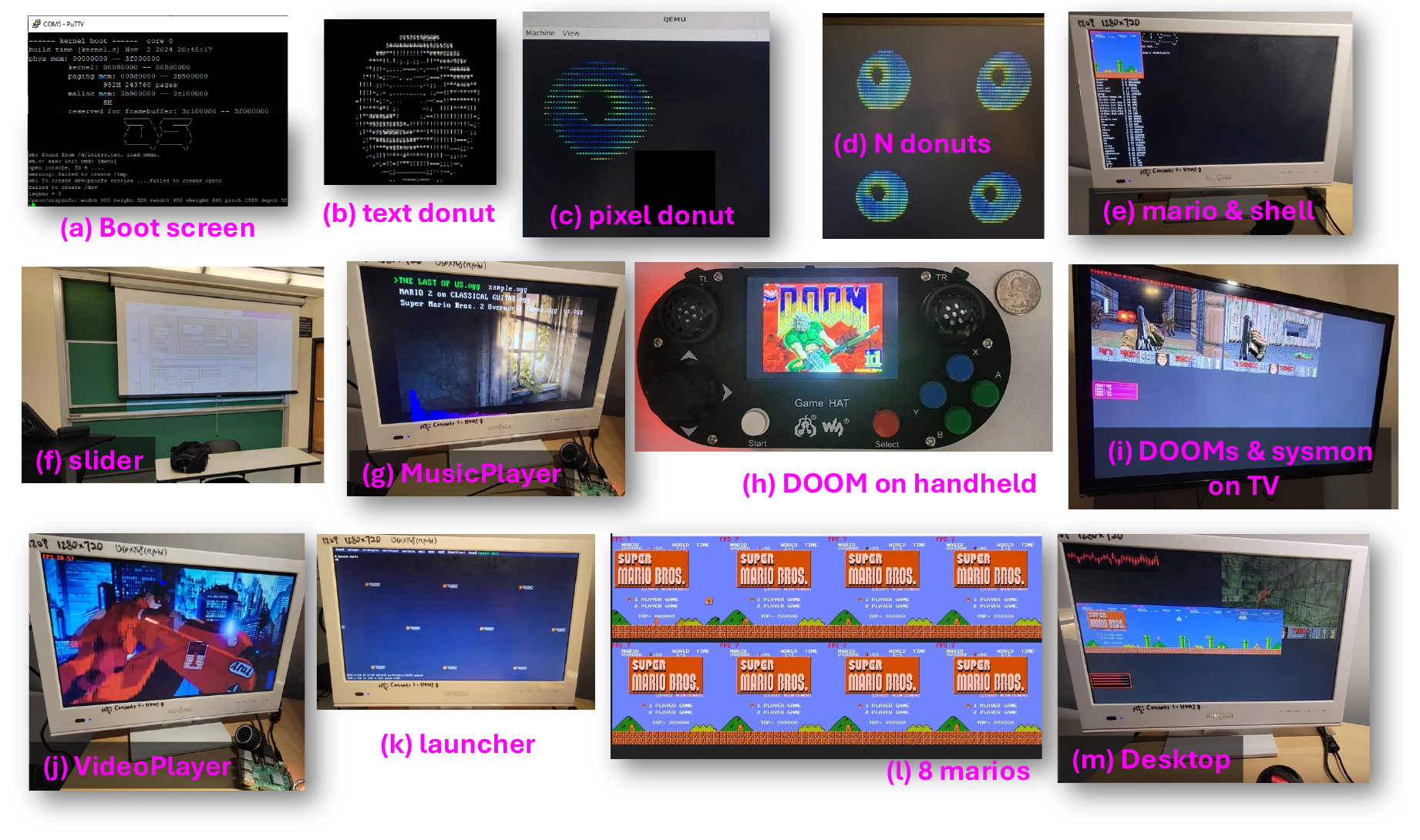}
		\captionof{figure}{\sys{} running various apps, across different prototype stages, and in different hardware settings. 
	(a)--(c) from Prototype 1, (d) from Prototype 2, (e) from Prototype 3, (f--g) from Prototype 4, (h--m) from Prototype 5.
	}
	\label{fig:collage}
\end{strip}

\begin{abstract}





    \sys{} is a new instructional OS that runs on commodity, portable hardware. It showcases modern features, including per-app address spaces, threading, commodity filesystems, USB, DMA, multicore support, self-hosted debugging, and a window manager. It supports rich applications such as 2D/3D games, music and video players, and a blockchain miner.
    Unlike traditional instructional systems, \sys{} emphasizes engaging, media-rich apps that go beyond basic terminal programs. 
    Our method breaks down a full-featured OS into a set of incremental, self-contained prototypes. Each prototype introduces a minimal set of OS mechanisms, driven by the needs of specific apps. The construction process 
    then progressively enables these apps by bringing up one mechanism at a time.

    \sys{} enables a wider audience to experience building a self-contained software system used in daily life.


\renewcommand{\thefootnote}{\fnsymbol{footnote}}
\footnotetext[1]{Co-primary authors.}
\renewcommand{\thefootnote}{\arabic{footnote}}

\end{abstract}

\ccsdesc[500]{Software and its engineering~Operating systems}
\maketitle

\section{Introduction}
\label{sec:intro}




Over the past several decades, the OS course has been a cornerstone of computer science education. While the specific objectives of OS courses may vary, they commonly aim to convey core system concepts: abstractions, virtualization, privilege separation, concurrency, and software/hardware interactions~\cite{ebling2024resources}.

Beyond these technical concepts, OS courses are also expected to offer a unique experience: building a working system that students can both demonstrate and use themselves. To this end, the OS course is the apex of an undergraduate CS curriculum, where students integrate their knowledge to construct a complete computer system.


Unfortunately, we observe a declining interest in OS---and systems software in general---among college students that we teach, owing to several factors:
the dominance of AI \cite{DominanceOfAI}, the perception that systems work is overly complex, and the frustration often associated with modern systems software. One telling indicator is that many premier North American universities ~\cite{csranking} (20 out of top 30) no longer require OS as a core course for undergraduate CS majors.
%
This shift presents both a \textit{wake-up call} and an \textit{opportunity} for OS educators:
(1) Instructional OS must stay relevant by appealing to broader students, highlighting its practical and demonstrable value.
(2) As an elective following \textit{Computer System Organization}, the OS course now addresses more motivated, better-prepared students. AI tools (e.g., Copilot~\cite{msft_copilot}) further empower them to take on complex code.

Today's OS education---and more importantly, the instructional OSes that support it---tend to prioritize teaching technical concepts~\cite{ebling2024resources}, while often overlooking the motivating power of real-world apps.
A unique strength of systems software, especially OSes, is that they can produce tangible artifacts that enable interesting, interactive apps. These artifacts can be demonstrated, played with, and even connected to students' other passions---whether in art, entertainment, design, or creative expression. This, in fact, aligns with the historical roots of many influential systems projects, which were originally created to serve personal or creative needs. Examples include UNIX, developed to run the game \textit{Space Travel}~\cite{salus1994quarter}; the World Wide Web~\cite{jacksi2019development}, created for sharing research information; and TeX, written for typesetting the author's own books~\cite{knuth1984texbook}.
Sadly, the spirit and joy of creating a usable system to support one’s own goals is often absent in today’s instructional OS projects.

\subsection{Principles}
Our response is an instructional OS called Proto, which manifests a guided journey through building an OS with rich applications.

Our first priority is to establish clear and compelling motivations for system building---motivations that go beyond running simple terminal programs.
To this end, we first target \textbf{(P1) appealing apps}: building engaging, interactive programs that ideally evoke an emotional response from the OS builders themselves. This motivation directly influences our design choices, such as treating the framebuffer as a first-class peripheral and prioritizing support for USB and audio IO. We further emphasize \textbf{(P2) demonstrability}. The OS should be able to run on real commodity hardware in addition to emulators. Ideally, it should be portable enough to carry around for live demonstrations, enabling students to showcase their work long after the course ends.

Our second consideration is managing complexity: balancing the technical depth needed to support engaging apps with the limited effort students can realistically invest (typically 17 weeks and no more than 150 hours). 
To achieve this, we adopt \textbf{(P3) incremental prototyping}. The OS is delivered to students in multiple self-contained snapshots, each augmenting earlier ones and is fully functional. 
Compared to a ``big bang'' approach, incremental prototypes provide clear baselines for continued development. 
We note that a number of prior instructional OSes were built with incremental
prototyping~\cite{JOS, chickadee, pfaff2009pintos}.
More broadly, incremental prototyping is a proven method in complex engineering projects such as the Ford Model T, Boeing 747, and SpaceX rockets~\cite{burchardt2020development, wu2012rapid, spaceX}. It is also common to many instructional OSes. We also require \textbf{(P4) minimum viable implementations}. Every new OS feature---whether it's a syscall, threading, or multicore scheduling---must directly support a target app or an essential component of it.
We will evaluate the efficacy of these principles in \sect{eval}.



\subsection{Design}
We first fully implement \sys{}, before decomposing it into five incremental snapshots, referred to as ``prototypes 1--5''.
Further, we map each prototype to one or multiple target applications, ranging from simple 3D animations (Prototype 1) to video games (Prototype 3) and video playback (Prototype 5). See Table~\ref{fig:features} for prototype features and their target apps. 

The OS construction by students
is then driven by enabling one app after another, augmenting earlier prototypes (so that they build upon familiar designs) and adding new OS features including
per-app address spaces, multimedia output, USB peripherals, filesystems, DMA,
and multicore execution.
At a higher level, this process resembles \textit{interactive designs}---akin to video games or theme park experiences---by keeping students as OS builders motivated with clear, meaningful goals, rewarding them with tangible progress (making apps functional), and maintaining engagement through timely transitions between different prototype stages in an incremental fashion.

\subsection{Results \& Contributions}
Through its design, \sys{} achieved the following goals. 

First, \sys{} is lightweight.
Its kernel core is less than 10K SLoC, comparable to existing instructional OSes (\S\ref{sec:eval:sloc}).

Second, \sys{} is demonstration-ready. 
It runs on both software emulators and low-cost ARMv8 hardware. 
As shown in Figure~\ref{fig:collage}, \sys{} supports a variety of apps: it can run games (e, i), play music and videos (g, j), and even project presentation slides for classroom use (f).

Third, \sys{} is fast and efficient. Through a careful balance of optimization and simplicity, \sys{} achieves strong performance on low-cost ARMv8 platforms. On kernel microbenchmarks (\S\ref{sec:eval:micro}), it incurs overheads similar to xv6~\cite{cox2011xv6};
it performs comparably to Linux and FreeBSD on most benchmarks. 
On app benchmarks (\S\ref{sec:eval:macro}), \sys{} runs \texttt{DOOM} at around 60 FPS and plays back video at around 26 FPS---on par with Linux and FreeBSD running the same apps on the same hardware. 
\sys{} also scales to quad-core processors.
When running on a handheld device (Figure~\ref{fig:collage}(h)), the system typically draws less than 4 watts and operates for several hours on a single battery charge (\S\ref{sec:eval:power}).

Finally, \sys{} is pedagogically effective. Our user study shows that most students found the design principles (P1--P4) directly contributed to their learning experience and sparked enthusiasm for systems software (\S\ref{sec:eval:user}).

In summary, this paper presents a series of incremental, self-contained OS prototypes motivated by appealing applications such as animations, media playback, and video games. 
Each prototype introduces a set of OS mechanisms explicitly mapped to target applications that critically depend on them.
We report the architectures chosen for each prototype, balancing trade-offs among app features, OS complexity, and modern OS concepts to be conveyed.
Our result is \sys{}, a new instructional OS that runs on commodity, portable hardware. 
It showcases modern features, including per-app address spaces, threading, commodity filesystems, USB, DMA, multicore, self-hosted debugging, and a window manager. 
It supports rich applications such as 2D/3D games, music and video players, and a blockchain miner.
The code is publicly available at: 
\url{https://github.com/fxlin/uva-os-main}. 

\section{Background}
\label{sec:motiv}
This section positions Proto within the broader context of instructional OSes and explains our rationale for hardware choices.




\subsection{Relation to existing OSes}
\label{subsec:existing_os}

\paragraph{Instructional OS}
Most instructional OSes are  	
``headless'', i.e., they lack GUI and instead support shell-based apps and utilities backed by simple file systems.
Notable academic projects include Embedded Xinu~\cite{gebhard2024using} (a port of Xinu~\cite{comer2015operating}), xv6~\cite{cox2011xv6}, Nachos~\cite{christopher1993nachos}, GeekOS~\cite{hovemeyer2004running}, Pintos~\cite{pfaff2009pintos}, MiniOS~\cite{otero2018minios}, and the more recent EGOS-2000~\cite{githubGitHubYhzhang0128egos2000}. The open-source community also contributes instructional OSes~\cite{osfs00, raspberry-pi-os}.
Among these, only a small subset can run on actual hardware, typically on ARMv7~\cite{francis2018raspberry,githubGitHubLdBECMXinu}, ARMv8~\cite{raspberry-pi-os,githubGitHubHongqinLirpios}, and RISC-V~\cite{gebhard2024using,cox2011xv6,githubGitHubYhzhang0128egos2000}.

While prior work---especially xv6---have greatly influenced our system, their designs heavily focus on conveying OS concepts, overlooking appealing apps (P1) and demonstrability (P2), which are central to our approach.


\paragraph{Production OS for education}
A recent survey~\cite{ebling2024resources} suggests that more than half of participating universities use Linux for teaching OS. Other choices of production OSes include Android~\cite{andrus2012teaching}, RT-thread~\cite{rt-thread}, and RISCOS~\cite{riscos}.
Students typically modify parts of these systems, such as CPU schedulers. 
This approach falls short for several reasons.
(1) It does not align with our goal of providing an end-to-end system-building experience. Modifying a large, existing OS does not offer the same educational value as building one from scratch.
(2) Production OSes are not suitable for stripping down due to their inherent complexity and generality. They support a wide range of apps and hardware, resulting in excessive layers of indirection.
(3) While working with production OSes like Linux/Android may excite some, their complexity may limit accessibility for a wider audience.

\paragraph{Full-stack system projects}
We are inspired by recent educational initiatives that guide students in constructing end-to-end systems capable of running graphical apps~\cite{one_student_one_chip,project-n}.
Compared to them:
(1) We target commodity hardware as the primary platform. This choice significantly influences how we balance practical realism and simplicity, handle hardware quirks, and support debugging (see Section~\ref{sec:overview}).
(2) We emphasize demonstration-ready, usable apps, which in turn motivate support for modern OS features such as commodity filesystems, USB, and DMA.

\subsection{Student assignments}
Next, we discuss student assignments built on prior educational OSes, including xv6~\cite{cox2011xv6}, Pintos~\cite{pfaff2009pintos}, Embedded Xinu~\cite{embedded_xinu_project}, GeekOS~\cite{hovemeyer2004running}, EGOS-2000~\cite{githubGitHubYhzhang0128egos2000}, Chickadee ~\cite{chickadee}, JOS~\cite{JOS}, and WeensyOS~\cite{CS_61}. Throughout, we use ``lab'' to refer to a single standalone submission from students. 


\paragraph{Lab structure and granularity}
Assignment design is largely shaped by the trade-off between learning objectives (breadth vs. depth) and the expected time commitment. Broadly, prior assignments fall into two types: breadth-oriented and depth-oriented. A breadth-oriented lab typically involves adding features to individual components~\cite{CS_61,hovemeyer2004running, black2009build, holland2002new}, whereas a depth-oriented lab often requires implementing most or all of an OS subsystem~\cite{pfaff2009pintos, JOS, chickadee}. Accordingly, breadth-oriented labs are usually independent (e.g., students receive fresh starter code for each lab), while depth-oriented labs are typically sequential, with each lab building on the codebase from previous labs~\cite{JOS, chickadee, pfaff2009pintos}, culminating in a functional OS by semester’s end.
Proto adopts a sequential approach in that labs build on each other, but is breadth-oriented in that each lab cuts across multiple OS components.

\paragraph{Lab topics}
Although labs vary in granularity, each typically centers on a single OS \textit{topic}---either a user component or a kernel subsystem. Common topics include:
(1) shell implementation~\cite{JOS,CS_61_pset5};
(2) CPU scheduling~\cite{otero2018minios,JOS,embedded_xinu_project,githubGitHubYhzhang0128egos2000};
(3) process management (e.g., \texttt{fork}/\texttt{exec}/\texttt{wait}/\texttt{exit})~\cite{cox2011xv6,chickadee,holland2002new,hovemeyer2004running};
(4) memory management (e.g., virtual memory or paging)~\cite{jos_project_list,CS_61_pset3};
(5) file systems (e.g., large-file support or fine-grained synchronization)~\cite{CS_61,chickadee,JOS,pfaff2009pintos}.
Few labs span user–kernel boundaries or integrate multiple subsystems.
In contrast, Proto’s labs are organized around target applications, so each lab naturally spans multiple subsystems and user–kernel interactions.

\paragraph{Student workload}
In prior educational OS courses, there are typically 4--7 labs per semester, each lasting 2--3 weeks. Labs are often divided into manageable checkpoints. Checkpoints usually have dependencies and are intended to be completed in sequence. Required submission artifacts are primarily source code and design documents.
Instructors typically provide starter code by removing key portions from a working implementation. The omissions range from a few essential functions (common in breadth-oriented labs~\cite{CS_61}) to most of an entire component (common in depth-oriented labs~\cite{pfaff2009pintos}).
Proto instead emphasizes breadth through application-driven labs of wider scope: students modify tens of files across multiple components, with tasks decomposed into smaller units and arranged in non-linear dependencies (see Section~\ref{sec:assignments}).

\subsection{Hardware Considerations}

Targeting real hardware aligns with our goal of demonstrability (P2); it exposes real-world challenges like cache timing, multicore bugs, and nondeterminism, which emulators cannot replicate.

We choose the Raspberry Pi 3B (Pi3) as our sole platform for several reasons. (1) \textit{Popular ISA}. ARMv8 is simpler than x86, yet widely used (IoT, phones, Macs), with extensive documentation and relevance. 
(2) \textit{Availability}. Released in 2016, Pi3 has sold over 23 million units, is priced at \$35, and will remain in production through 2029~\cite{tomshardware2024raspberry}. 
(3) \textit{Documentation}. Pi3 is well-documented, with strong open-source support and an active community~\cite{raspberrypiRaspberryRaspberry}. 
(4) \textit{Performance}. Four Cortex-A53 cores at 1 GHz, NEON SIMD, and 1 GB of memory make Pi3 suitable for a range of gaming and media apps. 
(5) \textit{Peripheral Ecosystem}. Pi3 is compatible with many expansion boards (called HATs~\cite{githubGitHubRaspberrypihats}), including a Game HAT (see Section~\ref{sec:eval}), which adds a 640p screen, battery, buttons, and speakers.

While other platforms based on RISC-V and x86
are also popular for instructional OSes~\cite{cox2011xv6,githubGitHubYhzhang0128egos2000}, 
they are lacking in one or more of the aspects above. 

\paragraph{Single Hardware Platform Support} 
Focusing on a single hardware platform simplifies both design and implementation.
Cross-platform compatibility would introduce extra layers of abstraction---such as function pointers and conditional compilation---that increase code complexity and reduce readability. It would also complicate the build system, making it harder to understand and modify. 

\paragraph{Risk of Hardware Obsolescence}
Targeting a single platform risks obsolescence if the hardware becomes unavailable. Historically, many x86-based instructional OSes faced this issue. We plan to switch to the Raspberry Pi 4 (also ARMv8) by 2030, updating the board-level code as needed.


\section{\sys{} Overview}
\label{sec:overview}

We next describe \sys{} in its complete form, which includes all the OS capabilities. 
This implementation is the starting point for our decomposition into a series of prototypes.

\paragraph{Userspace}
Our target apps include: 

\begin{myitemize}
    \item \texttt{donut}: One or more spinning 3D tori \cite{donut},  
    comprising either textual characters (for UART text output) or pixels (for framebuffer). 
    \item \texttt{mario}: The LiteNES emulator \cite{liteNes}, supporting limited games including Mario Bros. (1983). 
    \item \texttt{sysmon}: A floating, transparent window that visualizes real-time CPU and memory usage. 
    \item \texttt{slider}: A slide viewer for BMP, PNG, and GIF formats,
    intended for the OS builders to present their design.
    \item \texttt{DOOM}: A famous 3D game ported to virtually 
    ``anything with a screen'' \cite{itrunsdoom}. We build on the doom generic ~\cite{doom_generic} project and implement key event polling and texture/renderer management.
    We chose not to implement sound mixing due to its complexity.
    \item \texttt{MusicPlayer}: Intended to enhance user engagement and evoke emotions.
    It plays OGG files while displaying album covers. 
    The importance of sound in games/computer programs and for tangible user experience is crucial \cite{guillen2021role, kenwright2020there, robb2017impact}.
    \item \texttt{VideoPlayer}: Playback of MPEG-1 videos. 
    \item \texttt{launcher}: A GUI frontend for launching programs with an animated background. 
    \item \texttt{blockchain}: A multithreaded program for mining blocks. 
\end{myitemize}

Of these apps, a number of them were derived from NJU's project-N \cite{project-n}.
We also ported all console apps from xv6, including \texttt{shell} (enhanced with script execution) and utilities as \texttt{ls}, \texttt{cat}, and \texttt{echo}. 
Most user apps are in C and only a few are in C++ (\texttt{blockchain}). 

Underneath these apps are a small set of libraries we ported, 
including \texttt{libc} (\texttt{newlib}), \texttt{SDL} (Simple DirectMedia Layer, for portable rendering/input support), \texttt{libvorbis} (for \texttt{OGG} playback), \texttt{LODE} (for \texttt{png}), among others. 

\paragraph{OS architecture}
\sys{} is a monolithic kernel with an architecture similar to xv6.
The kernel and user apps run at \texttt{EL1} and \texttt{EL0} of ARMv8, respectively. The OS supports user processes and both user and kernel threads, and can execute them across all four CPU cores of the Pi3. The kernel is written in a mix of C and ARMv8 assembly. It includes CPU-specific code for the ARM Cortex-A53 and board-specific device drivers for the Pi3.

\sys{} uses virtual memory with separate address spaces for each user app. User space starts at \texttt{0x0}; kernel space uses addresses prefixed with \texttt{0xffff}. The kernel maps physical memory and I/O in 1 MB blocks, while user apps use 4 KB pages. Only the user stack uses demand paging.

\sys{} implements 28 syscalls across three categories: task management (e.g., \texttt{fork}, \texttt{sleep}), file system (e.g., \texttt{open}, \texttt{close}), and threading/synchronization (e.g., \texttt{clone}, \texttt{semaphore}).
The kernel exports device files (e.g., \texttt{/dev/fb} for framebuffer and \texttt{/dev/events} for keyboard) and proc files (e.g., \texttt{/proc/cpuinfo} and \texttt{/proc/meminfo}).
We chose UNIX-like kernel interfaces so that existing apps and their dependent libraries (e.g., \texttt{DOOM} and SDL) can be ported with minimal modifications.

\paragraph{IO support}
\sys{} supports a selective set of IO peripherals, tailored for the target apps: 
framebuffer (for HDMI output), USB keyboards, GPIO (for Game HAT buttons), 
PWM (for sound output via a 3.5mm jack), SD cards, and UART.
The decision to include USB is a careful balance between demonstrability and code simplicity, as discussed in Section~\ref{sec:design:proto4}.

\sys{} implements a window manager (see a photo in Figure~\ref{fig:collage}(m)) allowing multiple apps to render to the screen and for input events to be dispatched to the app with focus.

\paragraph{{OS image}}
On the Pi3, \sys{} runs from an SD card with two partitions. 
Partition 1 contains the kernel image, which is loaded and launched by the Pi3 firmware.
The kernel image packs the kernel code/data as well as an opaque ramdisk dump. 
The ramdisk, which is formatted with an ext2-like filesystem, 
incorporates all the user programs as \texttt{ELF} executables. 
Partition 2, formatted as FAT32, stores user files such as videos and game assets. 
At boot time, the kernel mounts its primary filesystem on the ramdisk 
and the secondary filesystem on the SD card's partition 2. 
This setup allows users to conveniently exchange files between \sys{} and their personal devices (computers, phones, cameras) by copying files to and from partition 2.
See \sect{design:proto5} for details and rationale for supporting FAT32.

\section{The prototypes}

\begin{table}[t!]
	\centering
	
	\includegraphics[page=1, width=.48\textwidth]{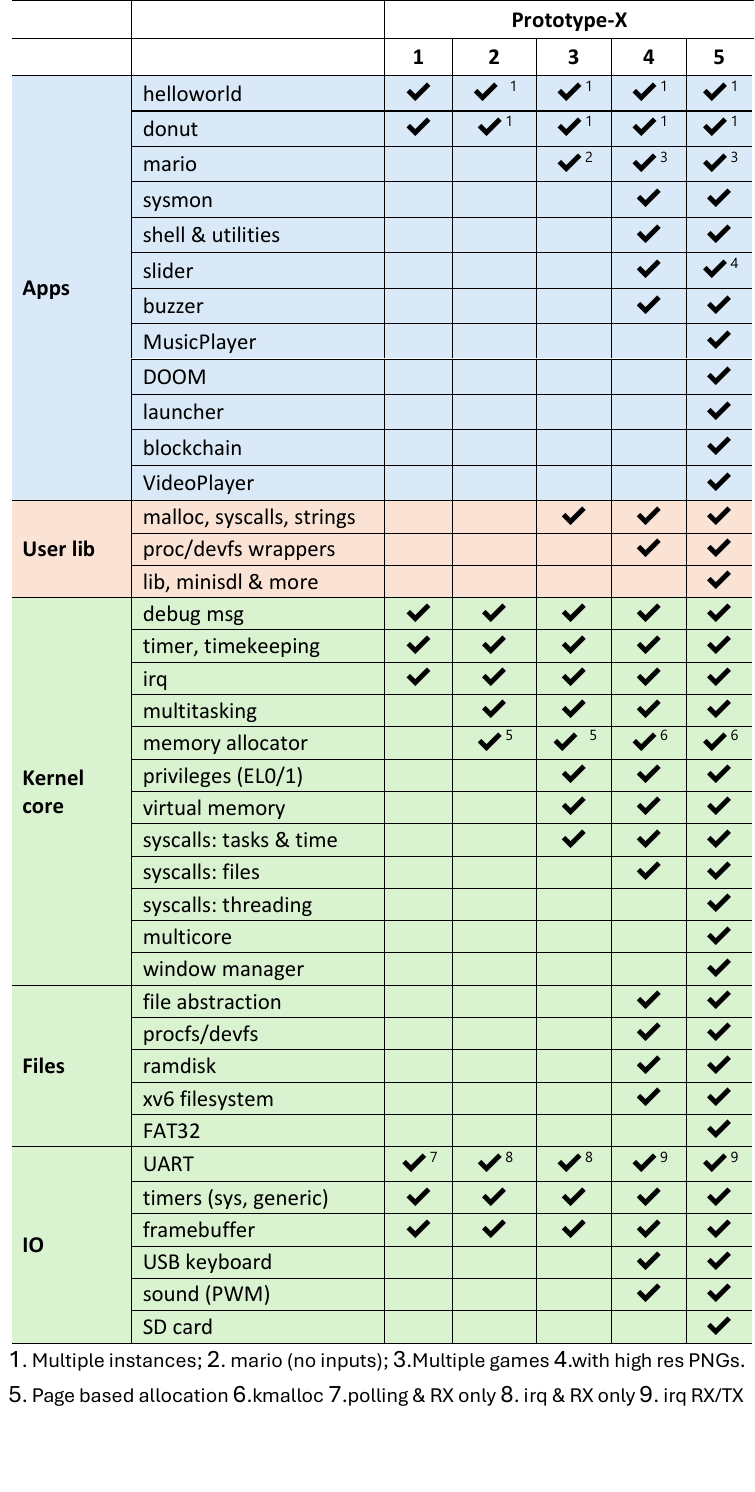}
    \caption{The feature matrix of all prototypes.}
	\label{fig:features}
\end{table}
\label{sec:design}

We now describe the individual prototypes, focusing on their target apps, the OS capabilities these apps necessitate, 
and the associated design tradeoffs.

\subsection{Prototype 1: ``Baremetal IO''} 
\label{sec:design:proto1}


\begin{figure}[H]
	\centering
	\includegraphics[width=0.48\textwidth]{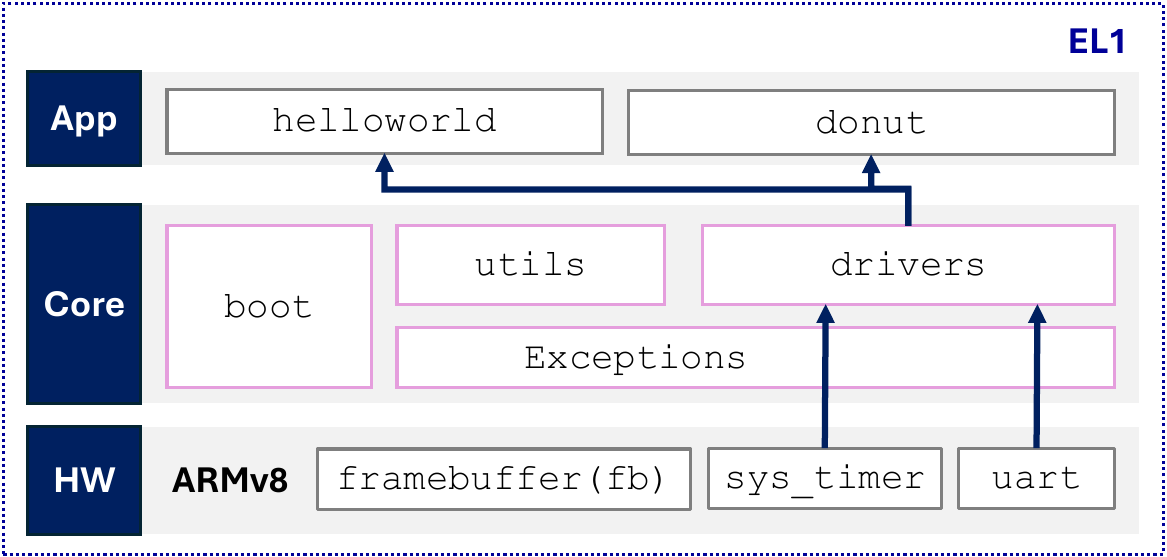}
	\caption{The architecture of Prototype 1}
	\label{fig:proto1}
\end{figure}

\textbf{Target app: } A donut spinning on display (Figure~\ref{fig:collage}(b,c)). 
\textbf{OS capabilities}: 
Prototype 1 is a baremetal \textit{appliance} for a single application. 
Its architecture is similar to that of a microcontroller like \texttt{ESP32}. 
It demonstrates the concept of software abstracting hardware details and introduces the idea of virtualization through the timer driver.

Unlike many education OSes where display was introduced much later if at all, we support the \textbf{framebuffer} as a first-class IO device via the Pi3's hardware mailbox interfaces. 
For debugging messages, the prototype implements a minimalist 
\textbf{output-only, polling-based UART} driver. 
For simplicity, UART writes are synchronous. 
We found interrupt-driven UART writes unnecessarily complicated the kernel;  
for instance, such writes would need a ring buffer for outstanding writes, which itself needs locks; 
on the other hand, the lock code may print debugging messages via UART, creating a circular dependency. 
Thus, we keep UART writes always synchronous throughout prototypes 1--5,  
and defer interrupt-driven (asynchronous) inputs and outputs to USB keyboards (Prototype 4) and DMA sound (Prototype 5), respectively. Finally, 
Prototype 1 drives per-core ARM \textbf{generic timers} and a SoC-level timer, as well as multiple virtual timers above these. 

Other than that, the kernel itself is minimalistic.
The prototype has output (for animation and debugging) but no input. 
Everything runs at the same exception level on a single core. 
For donut animation, each frame is rendered in the timer interrupt handler.
Synchronization is also introduced: initially a spinlock, which is later reduced to reference-counted interrupt on/off due to the single-core nature.

\subsection{Prototype 2: ``Multitasking''} 
\label{sec:design:proto2}


\begin{figure}[h]
	\centering
	\includegraphics[width=0.48\textwidth]{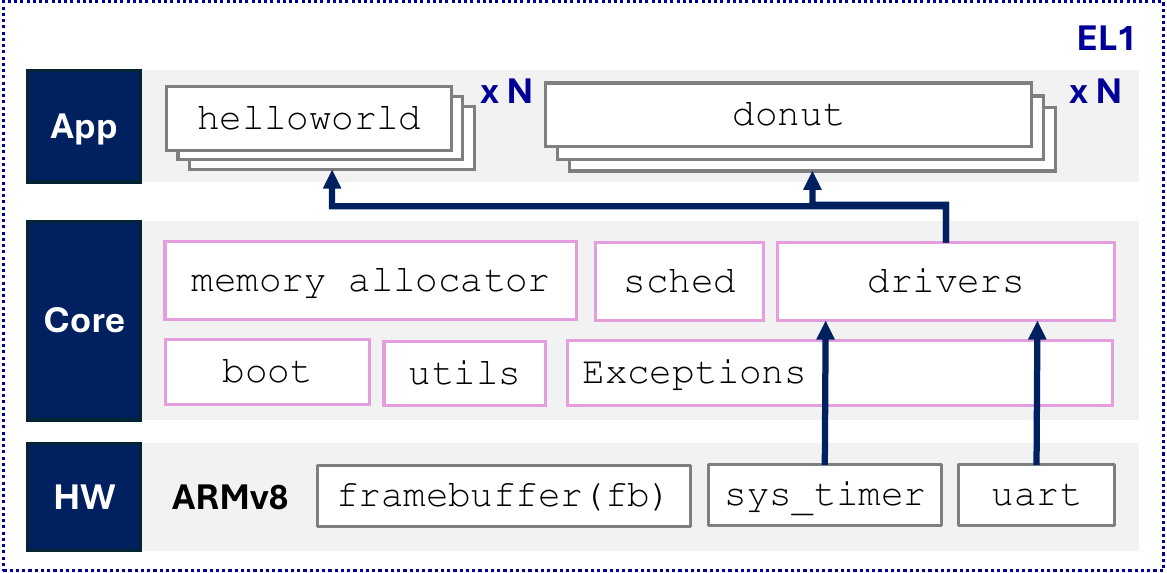}
	\caption{The architecture of Prototype 2}
	\label{fig:proto2}
\end{figure}

\noindent
\textbf{Target app}: Multiple concurrent donuts spinning at their own pace.
\textbf{New OS capabilities}:
This prototype introduces multitasking, runs predefined tasks, 
one for a donut. 
It reinforces the concept of virtualization, with the case of CPU.

Thanks to the framebuffer implementation in Prototype 1, 
the progress and priorities of tasks can be visualized on the screen 
through the spinning rates and angles of the donuts.
As timed animation, the donut tasks sleep periodically. 
This necessitates the scheduler's support for tasks to sleep. 
When all tasks are sleeping, the kernel makes the CPU idle, 
which motivates simple power management via \texttt{WFI} \cite{wang2023multiprocessing}.

Much of the complexity of enabling the target app comes from context switch: manipulation of CPU registers around the interrupt handlers and scheduler. 
The scheduler itself is simple (a single runqueue), sufficient to manage several tasks on a single core. 
All code runs at \texttt{EL1} with no kernel/user separation. 
Everything operates on physical addresses, with memory for tasks statically reserved. 
The ``kernel'' is a monolithic program, and apps are simply functions compiled into the same binary.

\subsection{Prototype 3: ``User vs. Kernel''} 
\label{sec:design:proto3}


\begin{figure}[h]
	\centering
	\includegraphics[width=0.48\textwidth]{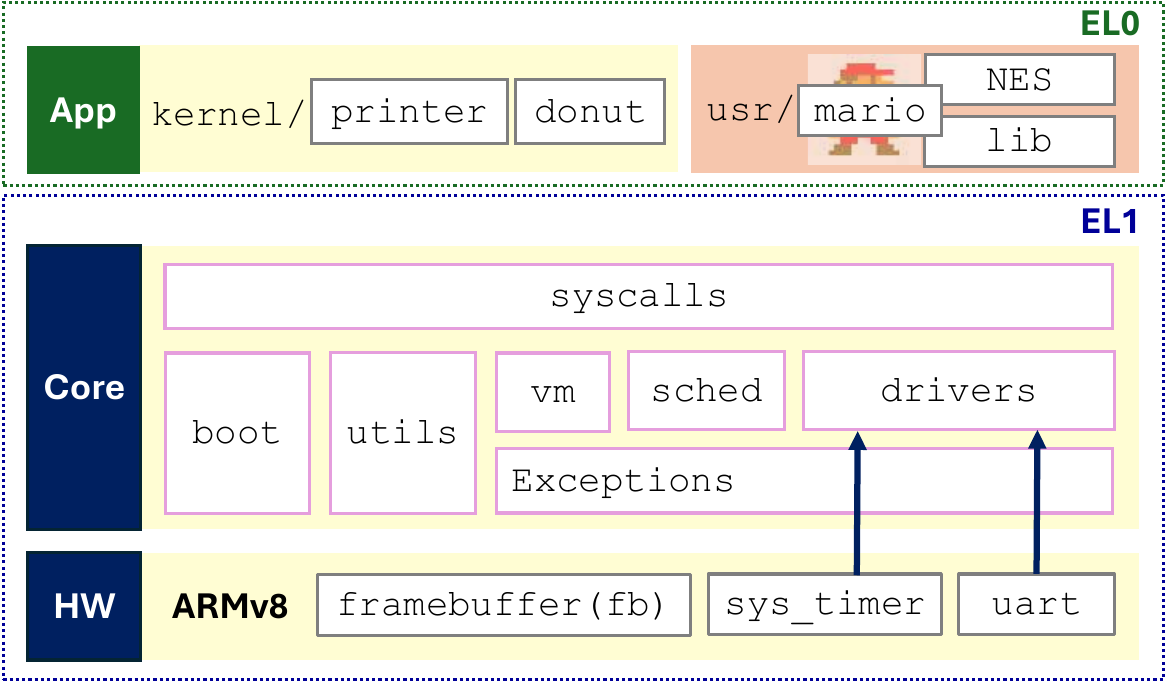}
	\caption{The architecture of Prototype 3}
	\label{fig:proto3}
\end{figure}



\noindent
\textbf{Target app}: \texttt{mario} without inputs (see \sect{overview} for description). 
\textbf{New OS capabilities}: 
This prototype introduces user/kernel separation, 
which allows us to compile, link, and execute \texttt{mario} independent of the kernel.


Our \textbf{virtual memory} system is standard. Prototype 3 enables virtual memory shortly after boot, using a small page table with coarse blocks to map 1GB of DRAM and IO registers linearly to the kernel's virtual address space. Each task maps a code/data region and a stack at 4KB granularity. Demand paging is enabled by default, initially mapping only code pages and one stack page. Tasks with repeated page faults at the same address are terminated by the kernel.

The VM is not enough for motivating the user-kernel separation.
To motivate a userspace more decoupled from the kernel, 
this prototype supports ``infant'' apps that run in their own address space and at \texttt{EL0}, 
but are still compiled and linked as part of the kernel. 
These infant apps are launched via jumps from the kernel code, rather than \texttt{exec()}. 
This is achieved by ensuring the app code only issues PC-relative memory references.
Evidently, such infant apps are brittle because absolute addresses would refer to the kernel address space and cause exceptions. They are only suitable for simple ``hello world'' code consisting of tens of lines. 
This limitation motivates linking and loading \texttt{mario} (around 2K lines of code) independently of the kernel. 

One important decision is the \textbf{omission of files}, deferring them to the next prototype, 
as file support would introduce a myriad of data structures (e.g. locks and inodes) and code, complicating the task management and syscall paths. 
Without files, our \textbf{syscalls} are just several: 
\texttt{fork()} and \texttt{exit()} for task lifecycles, 
\texttt{sbrk()} for \texttt{mario}'s pixel buffer allocation, 
\texttt{sleep()} for timed rendering, and \texttt{write()} hardwired to UART for debugging. 
While some have questioned the relevance of \texttt{fork()} in production systems~\cite{10.1145/3317550.3321435}, we found it valuable for exercising the concepts of VM in an educational context.
A \textbf{special file-less} \texttt{exec()}
is needed due to lack of files. 
Our build scripts bundle \texttt{mario}'s ELF executable as an opaque binary, with the kernel image at the link time.  
At run time, \texttt{exec()} parses the ELF region in memory and loads the code/data segments into the user virtual address space. It also hardcodes the arguments (e.g., framebuffer address and geometry) as expected by \texttt{mario}. 
 
At the end of \texttt{exec()}, the kernel appends a 4KB granular mapping to \texttt{mario}'s page table, covering the entire framebuffer (typically several MBs). For debugging ease, the virtual address is identity-mapped to the framebuffer's physical address, enabling direct framebuffer access akin to Linux's Direct Rendering Infrastructure (DRI) \cite{paul2000introduction}.
Related to the omission of files, \texttt{mario}'s input functionality is \textbf{deferred} to Prototype 4, 
which drives a USB keyboard and encapsulates it as a device file. 
Note that UART, although functional since Prototype 1, is unsuitable for games due to its lack of key modifiers, multi-key support, and key release detection. 

Interestingly, \texttt{mario}, even without inputs, continuously renders animated frames---such as flashing a coin on the title screen---and eventually transitioning into autoplay. 
This limitation encourages students to engage with the next prototype to enable full control of \texttt{mario}.

Framebuffer rendering provides a window into the impact of CPU cache behavior.
First, the framebuffer must be mapped as cached to avoid significant FPS drop. 
Second, an interesting yet subtle issue arises: the CPU cache must be flushed for the framebuffer region on every frame; 
otherwise, stale framebuffer pixels from previous frames will cause non-deterministic visual artifacts that will gradually disappear as cache lines hit the memory. 

Since the user level (\texttt{EL0}) cannot flush the cache under typical hardware configurations, a kernel request (e.g., via a syscall) is required. 
Ultimately, cache maintenance is crucial for reliable game performance.

Students work at both the user and kernel level to build Prototype 3. On the kernel side, they must implement key functions such as creating kernel page tables, setting up the kernel idmap, as well as completing the syscall path for \texttt{exec()}. On the user side, they are tasked with implementing argument parsing in the provided NES emulator and finishing the graphics buffer flushing function.



\subsection{Prototype 4: ``Files''} 
\label{sec:design:proto4}


\begin{figure}[h]
	\centering
	\includegraphics[width=0.48\textwidth]{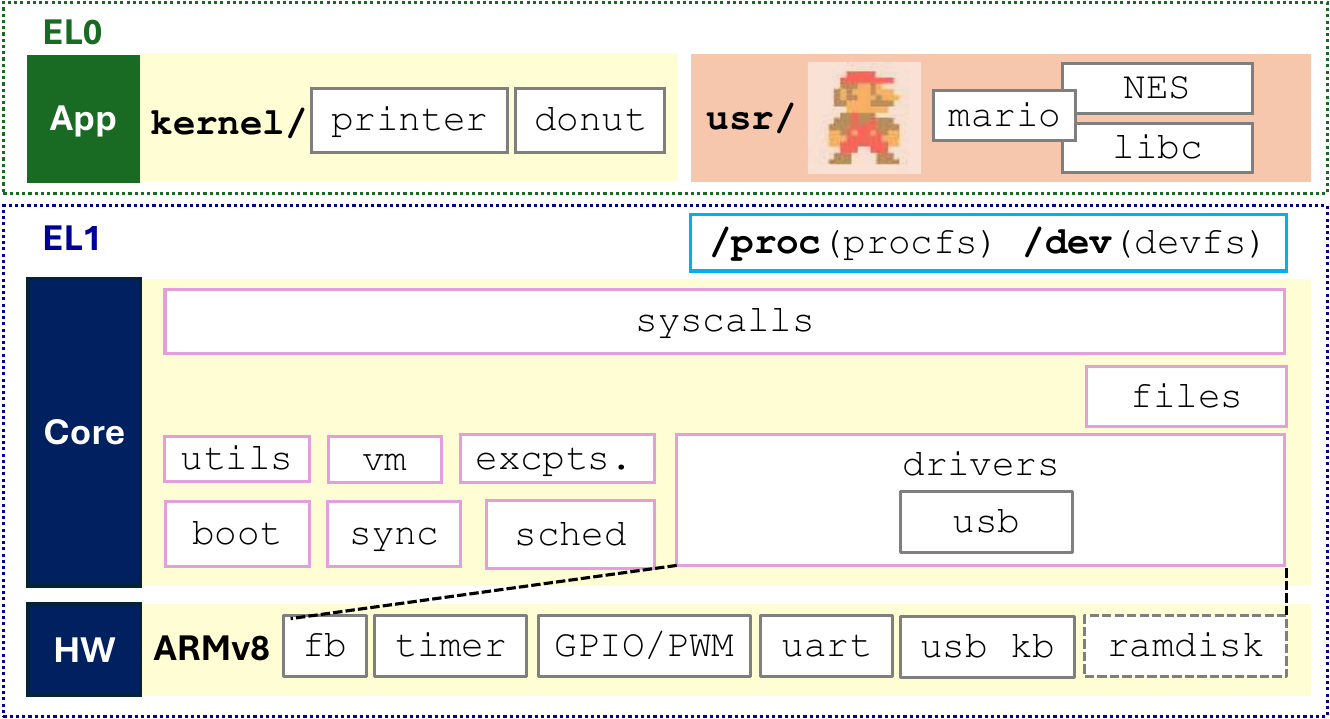}
	\caption{The architecture of Prototype 4}
	\label{fig:proto4}
\end{figure}

\noindent
\textbf{Target apps:} 
Additional NES games, slider, MusicPlayer. 
\textbf{New OS capabilities}: 
This prototype introduces the file abstraction, 
extending the user/kernel interfaces. 
Underneath files, it introduces a small filesystem on ramdisk as well as new IO drivers. 
The asynchrony introduced by the new IOs necessitates additional concurrency mechanisms inside both the kernel and userspace.


We store games, photos and music as disk files.
As such, the NES game engine can load \textbf{additional ROMs as files} (as opposed to the \texttt{mario} ROM hardcoded in a C header for the engine). 
The slider app loads \textbf{photo files} as slides, and MusicPlayer loads \textbf{music files} and their album cover photos. 
We ported xv6's ext2-like filesystem (``\texttt{xv6fs}'' for short) for its simplicity. 
The filesystem runs on a ramdisk; since all block reads/writes are synchronous, file read/write paths always execute in syscall contexts, easing understanding and debugging. 
Support for physical block devices, which must cope with storage hardware details, is deferred to Prototype 5. 
This decision decouples the complexity of a storage stack.

App IO is supported through \textbf{device files}.
They encapsulate IO devices, including the framebuffer (\texttt{/dev/fb}), audio output (\texttt{/dev/sb}), and USB keyboard (\texttt{/dev/events}), as needed by the target apps. Detailed descriptions are provided below. 

Atop disk and device files, Prototype 4 implements a series of file system syscalls, including \texttt{open()}, \texttt{close()}, \texttt{read()}, \texttt{write()}, and \texttt{lseek()}.

For audio playback, we implement \textbf{asynchronous IO paths}. 
The MusicPlayer app continuously writes sound samples to the device file (\texttt{/dev/sb}). The device driver then copies these samples and transfers them to the audio hardware via DMA for pulse width modulation output. Subsequently, DMA interrupts notify the driver to supply more samples. This setup forms a producer-consumer pipeline involving the app, the device driver, and the audio hardware, passing samples via ring buffers and synchronized by condition variables and locks—a classic OS design pattern. 
If there are problems in this pipeline—such as buffer corruption or deadlocks—audio playback will stutter or stop, providing immediate and clear feedback for debugging.


Compared to simple character devices such as UART, 
a keyboard that supports key modifiers is more suitable for games. 
We chose \textbf{USB keyboards} over simple I2C/SPI keypads as a deliberate tradeoff: 
USB keyboards are cheap (\$10) and available in portable or even foldable forms, easing system demonstration. 
The tradeoff is the increased software complexity from a USB stack.

We ported \texttt{Circle/USPi}~\cite{githubGitHubRsta2uspi},
a baremetal USB stack for Pi3. 
The stack only requires a few kernel APIs for virtual timers, which were introduced in Prototype 1. 
Although \texttt{USPi} is the lightest USB stack we could find, it still has around 10K SLoC, with multiple layers of abstraction.
We do not expect the students to fully understand and debug it. 
In return, the stack makes \sys{} extensible to more USB classes, such as ethernet adapters and mass storage, in the future. 

The app's need for handling multiple asynchronous inputs motivates \textbf{adopting IPC} for the app's event loop . 
Notably, the main loop of \texttt{mario} has to handle periodic timer events (for rendering frames at certain rate) and keyboard inputs. 
Lacking asynchronous IO and threading (available in Prototype 5),
\texttt{mario} shows a use case of IPC: 
the main loop forks two additional processes, one calling \texttt{msleep()} periodically and the other reading from \texttt{/dev/event} in a blocking manner; 
the two processes write events to a shared pipe (two writers), from which the main loop reads.

In this prototype, students work on both the kernel and user sides. On the kernel side, they implement functions that enable read and write operations on device files and complete the keyboard driver. On the user side, they develop code for terminal programs such as \texttt{shell}, \texttt{ls}, and \texttt{cat}, and write code to invoke system calls for reading and writing device files, thereby utilizing the framebuffer and keyboard.


\begin{figure}[h]
	\centering
	\includegraphics[width=0.48\textwidth]{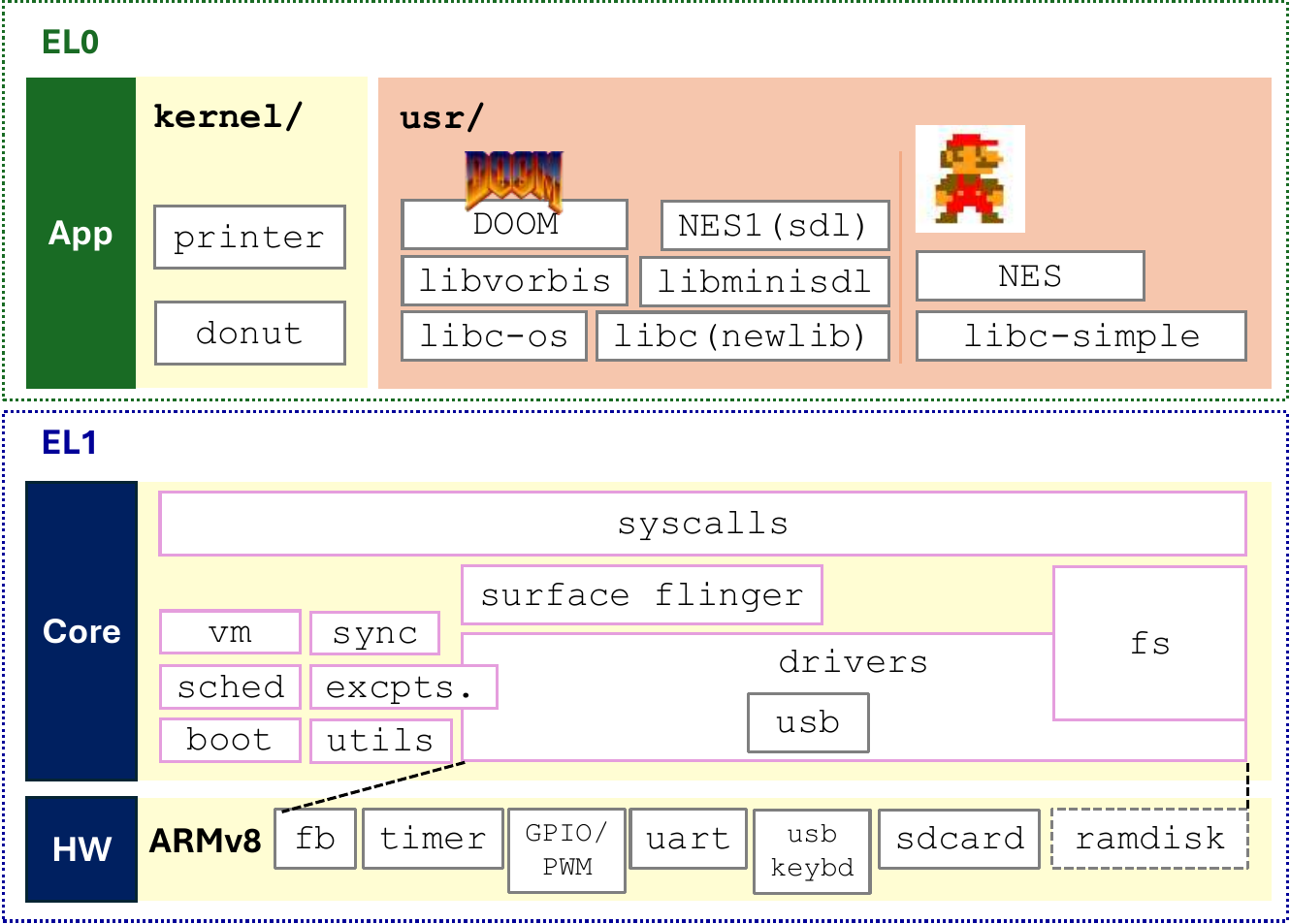}
	\caption{The architecture of Prototype 5}
	\label{fig:proto2}
\end{figure}
\vspace{-5mm}
\subsection{Prototype 5: ``Desktop''} 
\label{sec:design:proto5}

\paragraph{Target apps:} VideoPlayer, a blockchain miner, and DOOM, as well as running multiple instances of them concurrently on screen.
\textbf{New OS capabilities}: 
In the userspace, richer apps necessitate the inclusion of \texttt{libc} (newlib \cite{newlib})
and a trimmed-down SDL (Simple DirectMedia Layer) \cite{SDL} 
among other libraries.
The kernel brings up FAT32 on SD cards, 
scales to all four CPU cores, 
and runs a window manager. 

xv6's filesystem (``\texttt{xv6fs}'') becomes the bottleneck for new apps.
(1) Capacity: \texttt{xv6fs} only supports files up to 270KB, insufficient for larger photos, videos, and DOOM's game assets. 
(2) Performance: \texttt{xv6fs} reads and writes single blocks instead of block ranges, making it unacceptably slow for loading files in MBs.
(3) Interoperability: While it is possible to mount \texttt{xv6fs} on SD cards, 
commodity OSes cannot access these partitions without custom programs to parse and manipulate \texttt{xv6fs}.



These limitations motivate \textbf{porting a commodity filesystem} to Prototype 5.
We chose ChaN's \texttt{FatFS}, a FAT32 implementation used by popular embedded systems \cite{chan_fatfs}.
In around 6K SLoC, \texttt{FatFS} exposes high-level APIs with semantics close to our file syscalls (e.g. \texttt{open, close, read/write}). 
\texttt{FatFS} does not have an inode concept, encapsulating on-disk data structures behind its APIs, 
whereas each file or directory in \texttt{xv6fs} has an in-memory inode. 
To bridge the gap, Prototype 5 maintains \textit{pseudo}-inodes for each opened FAT file and directory. Prototype 5 further interposes file syscalls and dispatches them to either \texttt{xv6fs} or \texttt{FatFS} based on paths. At runtime, the OS mounts on its root filesystem (in \texttt{xv6fs}) under \texttt{`/'} and mounts the FAT32 partition under \texttt{`/d'}.


To support  \texttt{FatFS}, Prototype 5 includes a baremetal SD card driver. 
Unlike a full MMC (multimedia card) subsystem, which typically has tens of KSLoC \cite{mmc}, our driver is only 600 SLoC. It initializes the controller and card, and performs synchronous reads and writes of single blocks or block ranges, polling for their completion. Although the hardware is capable of DMA, our driver does not use it. 

We adopt \textbf{Non-blocking IO} for key-polling games, such as DOOM. 
This necessitates minor enhancements to the kernel IO path, including the VFS, device files, and the USB driver, to handle the non-blocking flag and peek the ring buffer without waiting for new data.

We use \textbf{threading} to process SDL audio concurrently. 
SDL uses a dedicated thread to stream audio samples to the device file, 
running in parallel with the thread decoding the audio.
Furthermore, structuring \texttt{mario} with multiple threads is more natural than using multiple processes that communicate via pipes (as in Prototype 4).
To this end, Prototype 5 implements a simple \texttt{clone()} with flag \texttt{CLONE\_VM} (a Linux-like interface chosen to ease the porting of existing user libraries);
within the kernel, threads are implemented by sharing \texttt{mm} structs across tasks.
To synchronize threads, Prototype 5 adds semaphore syscalls and implements user-level mutexes and condition variables atop semaphores. It also implements user-level spinlocks.



Running four or more \texttt{mario} instances quickly saturates a single core, motivating the OS to \textbf{scale to multicore}.
To do so, we made the following kernel modifications.
(1) The kernel's boot code releases additional cores held in the ``parked'' status by the Pi3 firmware.
(2) It configures MMUs for multicore cache coherency. 
(3) It augments the scheduler runqueue and vector tables so that each core has a copy. 
(4) The interrupts from ARM generic timers \cite{arm_generic_timer}, which drive scheduler ticks, are fed to each core,
whereas interrupts from all other IO are routed to core 0 for simplicity.


We implement \textbf{window manager} (WM, in around 800 SLoC) to provide desktop experiences.
Our window manager executes as a kernel thread.
Compared to running WM as a user process (e.g., as in Android), our design favors simplicity by avoiding the need to implement shared memory IPC and protocols for exchanging drawing contents across process boundaries.

To draw via the WM, apps render \textit{indirectly} to a device file (\texttt{/dev/surface}). 
Internally, WM maintains a list of such surfaces drawn by concurrent apps and periodically composites these surfaces onto the hardware framebuffer. 
Additionally, the WM tracks the ``focused'' app window and dispatches input events only to this app, which are returned to the app's \texttt{read()} of \texttt{/dev/event1}.
To display many apps simultaneously, WM can render overlapped windows and tracks their z-order. 
For efficiency, WM tracks dirty window regions across composition rounds and only redraws these regions.
It intercepts special key combinations (e.g., ctrl+tab) for switching window focus or moving windows around. 
It supports floating, semi-transparent windows, allowing \texttt{sysmon} to stay on top of all other running apps (see Figure~\ref{fig:collage}(m)).

In this prototype, students again work on both kernel and user code. On the kernel side, they implement system calls needed for the design, such as completing \texttt{clone()} for threading and modifying \texttt{open()} to support non-blocking I/O. They also finish the SD card driver and FAT filesystem to enable file system support. On the user side, students complete applications that leverage these new kernel features. For example, in the \texttt{MusicPlayer} app, they add a \texttt{clone()} call to create a worker thread that streams audio samples to the device file.



\section{Implementation}
\label{sec:impl}

\subsection{Debugging support}
Most instructional OSes rely on software emulators (e.g., QEMU) for debugging. 
However, this approach is inadequate to us, as certain nuanced and challenging bugs only manifest on actual hardware.
(1) CPU cache behavior. Real hardware may need manual cache flushes (e.g., for framebuffer updates) due to possible incoherence without proper MMU setup. 
(2) Uninitialized memory. Unlike QEMU's zeroed memory, real hardware has arbitrary values in uninitialized memory. 
(3) IO differences. E.g. GPU framebuffers may be mapped to arbitrary addresses on real hardware. 
(4) Parallelism. QEMU often lacks the non-deterministic thread interleaving seen on real hardware.
To address these issues, Proto supports debugging on both QEMU and real hardware.

In addition to UART-based \texttt{printk()} and emulator debugging, we support hardware-assisted debugging on Pi3:

\begin{myitemize}
    
    \item 
    \textbf{Hardware debug exceptions.} 
    We implement a simple debug monitor (200 LOC) using ARMv8 debug exceptions (\texttt{DBGBCR, DBGWCR}), enabling breakpoints (on specific PC), watchpoints (on address access), and single-stepping. This facilitates debugging complex user code and diagnosing memory corruption.
    
    \item 
    \textbf{Stack unwinder.}  
    We port a simplified ARMv8 stack tracer (600 LOC) \cite{stack_unwinder},
    capable of unwinding both kernel and user stacks. It prints raw callsite addresses (without debug symbols), which can be resolved offline on a development machine.

    \item 
    \textbf{Event tracing.}  
    Inspired by Linux's \texttt{ftrace}, we implement a ring buffer for all cores to write timestamped trace events, with negligible overhead. These are dumped on demand, e.g. to diagnose scheduler and concurrency issues.
    
    \item 
    \textbf{Panic button.}  
    We reserve ARMv8's FIQ (fast interrupt) for triggering emergency dumps. A GPIO-triggered FIQ (e.g., connected to a physical push button) remains unmasked at all times. When the button is pressed, the FIQ is routed in a round-robin fashion so that each occurrence is handled by a different core. 
    This mechanism ensures that if the kernel is deadlocked and IRQs are masked, pressing the button will dump call stacks from all cores.
\end{myitemize}

\paragraph{Consideration of JTAG and GDB}
Our kernel supports low-cost JTAG debuggers (e.g., FTDI-based, Segger J-Link) via the Pi3's GPIO pins. 
JTAG enables baremetal debugging from a development machine, typically using GDB over a proxy like OpenOCD~\cite{openOCD} (supporting breakpoints, single-stepping, etc).
However, our experience shows that JTAG on Pi3 requires tedious setup and lacks key features, such as halting on reset, loading programs, and reading system registers.
For our purposes, JTAG is less effective than self-hosted debugging described above.

Although  it is feasible to port GDB as a user program for native debugging, 
we decided against it due to the added dependencies, including \texttt{libreadline}, \texttt{libncurses}, kernel support for signals, and the \texttt{ptrace} syscall. 

\subsection{Performance Optimizations}
We implemented a small set of optimizations to make target applications usable.
Our kernel implements \textbf{memory move using ARMv8 assembly}, as these are often bottlenecks in framebuffer rendering. In the user library, we implement \textbf{pixel buffer conversions (e.g., YUV to RGB) using ARMv8 SIMD}. These optimizations improve video playback framerate by nearly 3x (from under 10 FPS to around 30 FPS for 480p video). 
Our kernel's buffer cache, inherited from xv6, supports only single-block operations, which suffices for xv6's simple filesystem but bottlenecks FAT32's multi-block access. This severely slows large file loads (e.g., DOOM assets, videos). To mitigate this, we 
\textbf{bypass the buffer cache for FAT32 range accesses}, directly interfacing with the SD card driver, reducing latency by 2--3x.
(See \sect{eval} for details).



\subsection{Support for C++ apps}
Our userspace implements a minimal runtime conforming to ARM's baremetal ABI (BPAPI ~\cite{armDocumentationx2013,osdevOSDevWiki}). This includes \texttt{crt0.c} (wrapping \texttt{main()}), and \texttt{crti.c/crtn.c} (handling global constructors/destructors). With under 100 SLoC, 
the runtime supports our C++ applications (e.g. Blockchain) as well as 
enabling future C++ apps like ML or game engines.

\subsection{Excluded OS Features}  
File crash-consistency was excluded due to the lack of compelling applications requiring crash consistency and the complexity of journaling mechanisms, which adds challenges to filesystem understanding.  
Network stack implementation, while feasible with our USB/Ethernet driver and available opensource TCP/IP stacks, was excluded as networking is not a course objective.  
Pthreads were omitted due to the lack of support in the recent \texttt{newlib} version and the complexity of popular implementations like \texttt{glibc} or \texttt{musl} ~\cite{musllibcMuslLibc}. Current parallel programs directly use syscalls for thread management.  
Signals and related syscalls were not implemented; this however limits our compatibility with tools like GDB.
HDMI audio output was excluded as it requires a proprietary firmware blob, adding unnecessary complexity for our purposes.



\subsection{Development and Deployment}
Leveraging the Pi3's rich peripheral ecosystem, we pair it with Game HAT, a handheld expansion board (Figure~\ref{fig:collage}(h)), which is more self-contained than other kits (e.g., RPi Personal Computer Kit, SmartiPi Touch). Game HAT's buttons connect via GPIO and emit key events through \texttt{/dev/events}. With this setup, \sys{} can render to a built-in 3.5" display as well as external displays such as big TVs (Figure~\ref{fig:collage}(i)) and classroom projectors (f). Section~\ref{sec:eval} evaluates battery life of \sys{} on Game HAT.

Our toolchain and scripts run on Ubuntu 22.04. To support Windows and macOS users, we test the setup on WSL2 (Windows), VMware Player (Windows/x86), and VMware Fusion (Apple Silicon).
We fixed a few QEMU bugs during development.

We develop \sys{} in three main stages:
(1) Implement all required features in a monolithic OS codebase;
(2) Identify target apps and map them to essential OS features; decompose the OS into five prototypes accordingly;
(3) Refine each prototype, define student experiments, and maintain them separately. We also introduce demo functionalities in the final prototype.





\section{Student assignments}
\label{sec:assignments}


\begin{table}[t!]
	\centering
	\small

	\begin{tabular}{lllll}
		\toprule
		\textbf{Lab} & \textbf{\#Tasks} & \textbf{\#Files} & \textbf{SLoC} & \textbf{\#Videos} \\
		\midrule
		Lab1 & 13 & 10 & $\sim$100 & 9 \\
		Lab2 & 10 & 10 & $\sim$100 & 9 \\
		Lab3 & 7 & 18 & $\sim$150 & 6 \\
		Lab4 & 7 & 21 & $\sim$300 & 7 \\
		Lab5 & 6 & 28 & $\sim$300 & 6 \\
		\bottomrule
	\end{tabular}
    \vspace{2mm}
    \caption{Student workload for labs, showing numbers of tasks, numbers of source files to modify, lines of code to write, and numbers of required video evidences for assessment. See Figure~\ref{fig:assignments} for detailed lab tasks.}
    \label{tab:effort}
\end{table}


\subsection{Lab design}
Throughout the semester, students complete five labs, each corresponding to one of the five prototypes described in Section~\ref{sec:design}.
For each lab, students submit one assignment.
We provide starter code for every lab, designed as a clean slate so that performance in an earlier lab does not affect grading in later labs.

\paragraph{Task structure}
Assuming students have at most 150 hours for assignments (\S\ref{sec:motiv}), we design labs to prioritize breadth of OS concepts over depth in any single component.
Our goal is for students to work with many parts of an OS, with each part requiring relatively small (yet still crucial) tweaks or fixes.
Each lab consists of multiple tasks, where each task exercises specific OS concepts, as shown in Figure~\ref{fig:assignments}.
Tasks often have dependencies, and completing a sequence of dependent tasks unlocks interesting application features along the way.
Students are not required to complete all tasks to earn full credit for a lab; 
they may choose their own path through the task graph based on the features they aim to implement.
Some students find this structure analogous to video games, with a main storyline complemented by optional side quests.

\paragraph{Task workload}
We, as instructors, create the starter code by removing key pieces from a working prototype.
Each task is intentionally small, typically requiring students to understand 1–2 source files, modify several or more code locations,
and write tens of lines of code in total.
For example, one task asks students to complete the user-level memory mapping code so that \texttt{mario} can render graphics.
Table~\ref{tab:effort} summarizes the typical student workload.
This contrasts with many instructional OS assignments that require implementing an entire standalone subsystem (e.g., a filesystem).

\paragraph{Teamwork}
Labs 1–3 are completed individually, while Labs 4 and 5 expect teamwork. 
The rationales are:
(1) Labs 4 and 5 involve rich userspace and GUI applications, which may exceed the time budget of some students working alone;
(2) many tasks in Labs 4 and 5 have inherent parallelism (e.g., separate kernel and user-level components needed by a window manager), allowing teammates to work concurrently and later integrate their work.

\subsection{Student deliverables \& assessments}
We find it challenging to automatically test a custom OS running in QEMU or on real hardware. 
Thankfully, our labs are centered on media-rich applications; 
we thus require \textit{video evidence} for critical tasks (marked with bold borders in Figure~\ref{fig:assignments}), making
students responsible for demonstrating that their OS enables the required features (e.g., rendering four rotating donuts at different rates).
Video submissions must meet strict criteria: minimum duration (e.g., 30 seconds), minimum resolution (e.g., 1080p), continuous display of the student's university ID, and no post-editing.
If video evidence is not provided, instructors will manually assess the student's source code.


\section{Evaluation}
\label{sec:eval}


We test our OS prototypes with apps on Raspberry Pi 3 (Pi3 in short) and QEMU as listed in ~\autoref{tab:platforms}.
The evaluation shows our system (1) is capable of running feature-rich apps, (2) is competitive against production OSes on target apps, not only just microbenchmarks, (3) is sufficiently efficient for demonstration and use.


\begin{table}[t!]
	\centering
    \small

	\begin{tabular}{ll}
	\toprule
	\textbf{Platform} & \textbf{Configuration} \\
	\hline
	Pi3 & Pi3 model b+, Samsung EVO MicroSD 32GB \\ \hline
	qemu-wsl & QEMU on Ubuntu in WSL2 on Win11 \\
	qemu-vm & QEMU on Ubuntu in VMPlayer on Win11 \\ \hline
	\bottomrule
	\end{tabular}
    
    \vspace{2mm}
    \caption{\textbf{Test platforms used in evaluation}. QEMU runs on a machine with Intel Ultra 7 155H and 96GB of DRAM}
    \label{tab:platforms}
\end{table}

\begin{table}[t!]
	\centering
    \small

	\begin{tabular}{lll}
		\toprule
		\textbf{OS} & \textbf{C Library} & \textbf{Media Library} \\
		\midrule
		Ours       & newlib 4.4.0     & (custom) \\
		xv6-armv8~\cite{githubGitHubHongqinLirpios}            & musl 1.2.1       & None \\
		Ubuntu/Linux 22.04    & glibc 2.35       & SDL 2.0.20 \\
		FreeBSD 14.2        & BSD libc 1.7   & SDL 2.30.10 \\
		\bottomrule
	\end{tabular}
    \vspace{2mm}
    \caption{\textbf{OS configurations used in evaluation}}
    \label{tab:os-config}
\end{table}


\subsection{Code analysis}
\label{sec:eval:sloc}

\begin{figure}[t!]
\begin{minipage}{0.22\textwidth}
	\centering
	\includegraphics[width=1\textwidth]{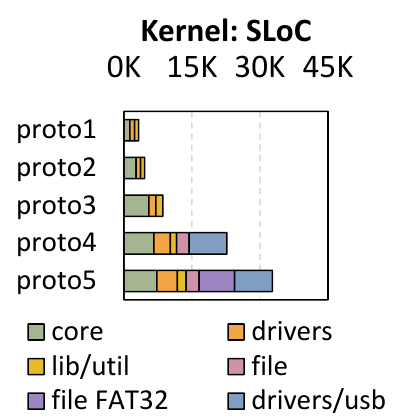}
\end{minipage}
\hfill
\begin{minipage}{0.24\textwidth}
    \centering
    \small
    \begin{tabular}{lp{2cm}}
        \toprule
        \textbf{Proto.} & \textbf{Details} \\
        \hline
        proto1 & app: $\sim$260 \\
        \hline
        proto2 & app: $\sim$557 \\
        \hline
        proto3 & app: $\sim$2.4K \\
        & userlib: 781 \\
        \hline
        proto4 & app: $\sim$5K 
        \\ & userlib: 957 \\
        \hline
        proto5 & app: $\sim$76K 
        \\ & userlib: $\sim$770K 
        \\
        \bottomrule
    \end{tabular}
\end{minipage}
\caption{\textbf{Source code analysis}. (Left) Breakdown of \textit{kernel} source lines of code (SLoC) for different prototypes; (Right) Breakdown of \textit{app} source lines of code (SLoC) for different prototypes.}
\label{fig:sloc}
\end{figure}

Figure~\ref{fig:sloc}(a) shows the kernel SLoC, ranging from approximately 2.5K (Prototype 1) to around 33K (Prototype 5, mostly from FAT32 and USB drivers), 
comparable to other instructional OSes (xv6 \cite{cox2011xv6}: 10K, Xinu \cite{comer2015operating}: 8K, pennOS \cite{aviv2012experiences}: 6K).
The breakdown highlights our incremental OS construction; the core functionalities remain small (from around 1K SLoC in Prototype 1 to around 8K in Prototype 5). 
Figure~\ref{fig:sloc}(b) shows the userspace source: ranging from simple animations in a few hundred SLoC to DOOM in 45K SLoC. Much complexity comes from a few production libraries such as \texttt{newlib}.


%

\begin{figure}[t!]
\begin{minipage}{0.21\textwidth}
    \centering
    \includegraphics[width=\textwidth]{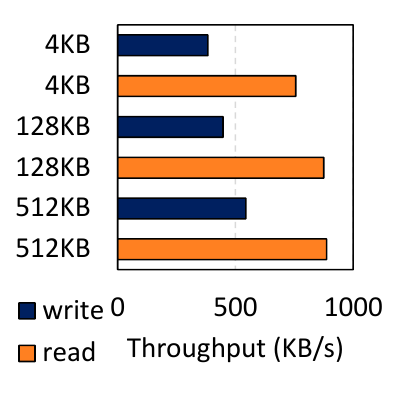}
\end{minipage}
\hfill
\begin{minipage}{0.24\textwidth}
    \small
    \begin{tabular}{lc}
        \toprule
        Syscall & 3.4±0.04 us \\
        IPC latency & 21.0±0.00 us \\
        kernel load \\ –by firmware & 2753±3 ms \\
        kernel boot \\ –to prompt  & 3186±7 ms \\
        \bottomrule
    \end{tabular}
\end{minipage}
\caption{\textbf{Kernel microbenchmarks}. (Left) File system throughput;
(Right) latencies in syscall (getpid()) and IPC (one-byte message sent over pipe()), averaged over 5,000 runs; boot time from power on to kernel loaded and to shell prompt, averaged over 5 runs. Platform: Pi3}
\label{fig:fs}
\end{figure}

\subsection{Microbenchmarks}
\label{sec:eval:micro}

The building blocks of our OS exhibit low overheads, as shown in \autoref{fig:fs}:
(1) The overhead of a syscall is around 3 $\mu$s.
(2) One-way IPC between user processes takes approximately 21 $\mu$s, which includes scheduling, data copying, and context switch; 
(3) The FAT32 read/write throughput achieves several hundred KBs per second; 
the throughput is primarily constrained by our SD driver, which lacks DMA support and polls registers for read and write.
(4) The entire device takes around 6 seconds from power-on to shell prompt; much of this time is spent by the firmware loading our kernel and our kernel initializing slow peripherals such as USB controllers.

\begin{figure}[t!]
	\centering
	\includegraphics[width=0.48\textwidth]{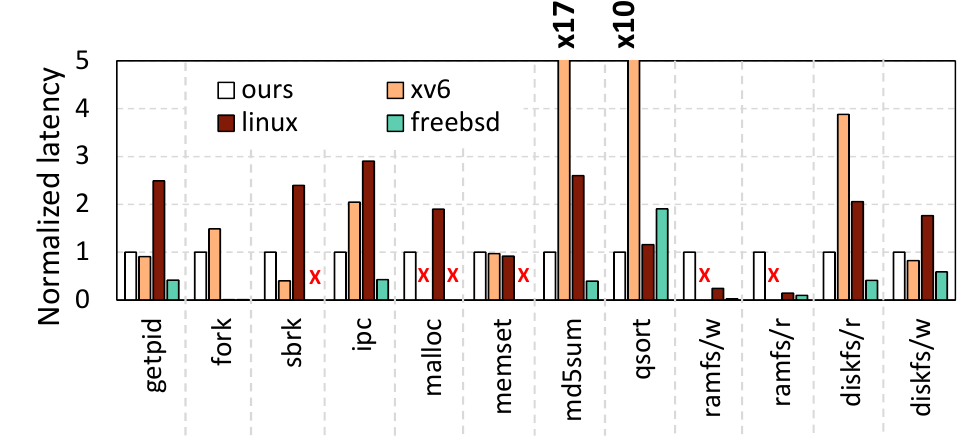}
	\caption{\textbf{OS microbenchmarks}, showing our latencies (normalized to 1.0) are lower than xv6 on most benchmarks, and are comparable to Linux and FreeBSD. 
	All experiments run on Pi3. 
	{\color{red}x} on empty bars indicates benchmarks that could not be run due to missing OS features.}
	\label{fig:micro-cmp}
\end{figure}

\paragraph{Comparison to xv6}
Figure~\ref{fig:micro-cmp} compares our system to the xv6-armv8~\cite{githubGitHubHongqinLirpios}, 
a popular instructional OS ported to Pi3, using \texttt{musl}~\cite{musllibcMuslLibc} as its C library. 
The results indicate: 
(1) Our system exhibits comparable kernel overheads, as reflected by similar latencies in benchmarks such as \texttt{getpid}, \texttt{sbrk}, and \texttt{IPC}. 
(2) Our system outperforms xv6-armv8 on compute benchmarks (e.g., \texttt{md5sum} and \texttt{qsort}), likely due to differences in the standard C libraries used (newlib in ours versus \texttt{musl} in xv6-armv8). 
(3) Our system also outperforms xv6-armv8 on file benchmarks, which can be attributed to differences in SD card driver implementations; both are simplistic and custom-built, but ours appears to be more efficient.

\paragraph{Comparison to production OSes}
Figure~\ref{fig:micro-cmp} compares our system to Ubuntu 22.04 and FreeBSD 14.2. The results indicate that our system achieves comparable latencies (within 0.5x to 2x) to production OSes in most cases. Notable exceptions are: 
(1) Our \texttt{fork()} (also xv6) is significantly slower due to the lack of lazy page table replication and demand paging as in production OSes. 
(2) Our file performance is also lower due to simplistic filesystem implementations and SD card drivers that use polling.

We also evaluated a few other instructional OSes that support Pi3, 
such as vmwOS ~\cite{francis2018raspberry,githubGitHubDeatervmwos}, 
Xinu~\cite{githubGitHubLdBECMXinu,comer2015operating}, and RT-Thread~\cite{rt-thread}. 
However, they were unsuitable for benchmark comparisons: some only support 32-bit execution, others are restricted to a single address space, and many lack sufficient system call implementations to run most of the microbenchmarks.



\subsection{App benchmarks} 
\label{sec:eval:macro}




\begin{figure*}[t!]
\begin{minipage}{0.64\textwidth}
	\centering
    
    \small
	\begin{tabular}{lccc|cc}
	\toprule
    \textbf{Platform} & \multicolumn{3}{c|}{\textbf{Pi3}} & \textbf{qemu-wsl} & \textbf{qemu-vm} \\
	& \textbf{Ours} & \textbf{Linux} & \textbf{FreeBSD} & \textbf{Ours} & \textbf{Ours} \\
	\cmidrule(lr){2-4} \cmidrule(lr){5-6}
	 \textbf{DOOM} & 61.80±0.01 & 31.88±0.13 & 51.24±0.00 & 99.86±0.50 & 92.13±0.70 \\ \hline
	 \textbf{video (480p)} & 26.68±0.40 & 19.00±0.20 & 24.40±0.13 & 30.26±0.04 & 28.18±0.28 \\ \hline
	 \textbf{video (720p)} & 11.57±0.25 & 10.05±0.12 & 14.60±0.21 & 18.37±0.83 & 15.91±0.47 \\ \hline
	 \textbf{mario-noinput} & 108.11±7.35 & - & - & 137.55±9.53 & 106.16±2.71 \\ \hline
	 \textbf{mario-proc} & 114.72±0.55 & - & - & 143.37±2.28 & 185.69±2.41 \\ \hline
	 \textbf{mario-sdl} & 72.20±0.59 & 87.28±5.39 & 56.38±0.50 & 121.55±2.48 & 192.98±1.55 \\
	\bottomrule
	\end{tabular}
    \vspace{2mm}
    \captionsetup{type=table} 
	\caption{
    \textbf{Throughput (FPS) of benchmark apps}. 
	See Table~\ref{tab:platforms} for platforms and Table~\ref{tab:os-config} for OS configurations.
	Linux/FreeBSD do not run \texttt{mario-noinput|proc} (marked as `-'), which depend \texttt{devfs} and \texttt{procfs} interfaces specific to our OS.	
	}
    \label{fig:throughput}
	
\end{minipage}
\hfill
\begin{minipage}{0.3\textwidth}
	\centering
	\includegraphics[width=0.9\textwidth]{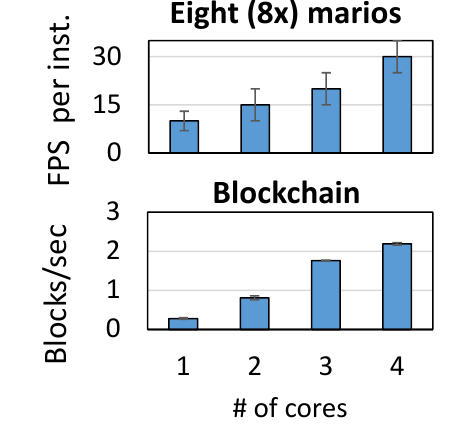}
	\caption{\textbf{FPS per app instance as a function of number of CPU cores}. Our OS scales to up to four cores. Platform: Pi3 with a quad-core CPU.}
	\label{fig:scalability}
\end{minipage}
\end{figure*}

\sys{} delivers good performance for its target apps.

\paragraph{Applications}
Our benchmark configurations include: 
(1) DOOM, which uses direct rendering without a window manager. 
(2) VideoPlayer, which decodes both video and audio streams and uses direct rendering; video files are preloaded into memory before decoding.
(3) \texttt{mario} in three variants. 
(i) \texttt{mario-noinput} (Prototype 3), which runs one thread and uses direct rendering, without handling any input events.
(ii) \texttt{mario-proc} (Prototype 4), which runs multiple processes to handle input events; it uses direct rendering. 
(iii) \texttt{mario-sdl} (Prototype 5), which runs multiple threads to handle the events; it renders indirectly via a window manager.

\paragraph{Throughput} (\autoref{fig:throughput})
In these experiments, we make DOOM and \texttt{mario} variants render as fast as possible without locking to a fixed FPS; we make video playback target a video's native framerate. 
FPS is measured after a 20-second warm-up period. 
As illustrated in ~\autoref{fig:throughput}, both DOOM and \texttt{mario} on \sys{} achieve decent throughput of 62--115 FPS. 
In video playback, our system attains 27 FPS for 480p video, close to its native framerate (30 FPS), while considerably lower for 720p video (12FPS). 
Among three \texttt{mario} variants,
\texttt{mario-noinput} has similar FPS to \texttt{mario-sdl}, which are both lower than \texttt{mario-proc}; 
the differences reflect the impact of our OS designs, which will be analyzed in latency breakdown below. 
All benchmarks see higher throughputs (from 13\% to 1.5x) when \sys{} runs on QEMU atop a modern x86 machine.

\paragraph{Comparison to production OSes}
We further compare our throughputs to Ubuntu/Linux and FreeBSD (configurations in Table~\ref{tab:os-config}). 
They run on the same hardware (Pi3), 
use the same compiler and optimization levels (\texttt{gcc-11} with O2),
and build the same app sources. 
For minimum overhead, we run Linux and FreeBSD on X servers without a window manager. 
The results show that our throughputs are comparable to Linux/FreeBSD, 
ranging from 0.8x to 1.9x, or differing by \(-15\)--30 FPS. 
Note that we do \textit{not} claim that our system is competitive against production OSes, 
as the latter is capable of hardware-accelerated rendering which our apps do not utilize; 
and that the latter incurs the overhead of a generic X server which our system eschews.

\begin{figure}[t!]
	\centering
	\includegraphics[width=0.35\textwidth]{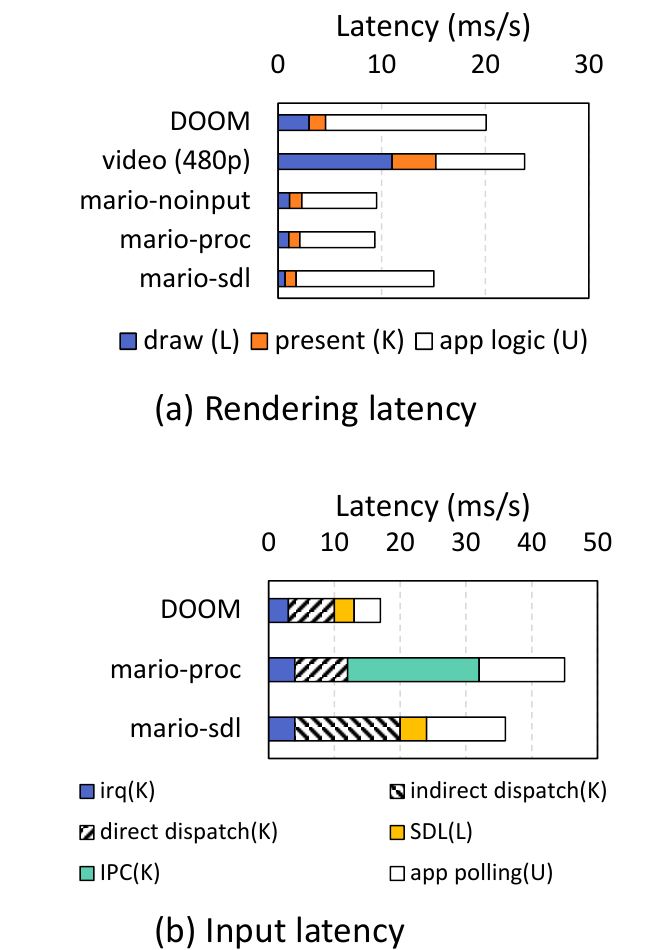}
	\caption{\textbf{Latency breakdown of benchmark apps}. Note that VideoPlayer and \texttt{mario-noinput} has no inputs. In legends: K=kernel, U=user app, L=user library}
	\label{fig:latency}
\end{figure}

\paragraph{Rendering latency}
Figure~\ref{fig:latency}(a) reveals the sources of rendering overhead. 
Overall, rendering delays are dominated by app execution: 
primarily the application logic such as game engine and video/audio decoding and drawing (e.g. buffer and pixel manipulation);
by contrast, the kernel incurs low overhead, e.g. writes to \texttt{mmap}'d framebuffer or device files.  
The latency also explains the FPS difference when running the three \texttt{mario} variants on our OS (\autoref{fig:throughput}): 
\texttt{mario-sdl} shows higher latency in its app logic than the other two variants,
likely because inclusion of a full-fledged C library (\texttt{newlib}) adds overhead to its game execution. 

\paragraph{Input latency}
We evaluated system responsiveness, 
by tracing a USB keyboard event traveling from the device driver to app. 
In the experiments, we cap the frame rate at 60 FPS. 

Overall, such end-to-end input latency corresponds to as low as 1–2 game frames. 
A detailed breakdown in Figure~\ref{fig:latency}(b) shows that: 
(1) the majority of the latency comes from the app logic: 
both DOOM and \texttt{mario} poll for input events in their main event loops, 
in which polling intervals (5 to 13 ms) contribute noticeable latency.  
(2) Among the remaining components, the extra event indirection (from window manager for \texttt{mario-sdl}, and the IPC between main/input tasks for \texttt{mario-proc}) contribute significantly, which is the cost for OS modularity. 
(3) Notably, our kernel's \texttt{pipe()}, a simplistic design ported from xv6, becomes a bottleneck even for passing a keyboard event (fewer than 10 bytes). 
DOOM, without such event indirection, shows the lowest latency. 


\paragraph{Multicore scalability} 
Our system scales well with the number of CPU cores, 
on a \textit{multi-programmed} workload (eight simultaneous \texttt{mario} instances, screenshot in Figure~\ref{fig:collage}(l)) 
as well as on a \textit{multi-threaded} workload (Blockchain miner).
As shown in ~\autoref{fig:scalability}, as the number of cores increases, 
the throughput sees proportional growth. 
We also verify that the utilization of all CPU cores is constantly over 95\%. 


\paragraph{Memory consumption}
\sys{} has low memory usage. When running target apps individually (e.g., \texttt{mario}, DOOM, and VideoPlayer), we measured total OS memory usage between 21 MB and 42 MB — about 2\% to 4\% of the Pi3's 1 GB of DRAM.

\begin{figure}[t!]
	\centering
	\includegraphics[width=0.48\textwidth]{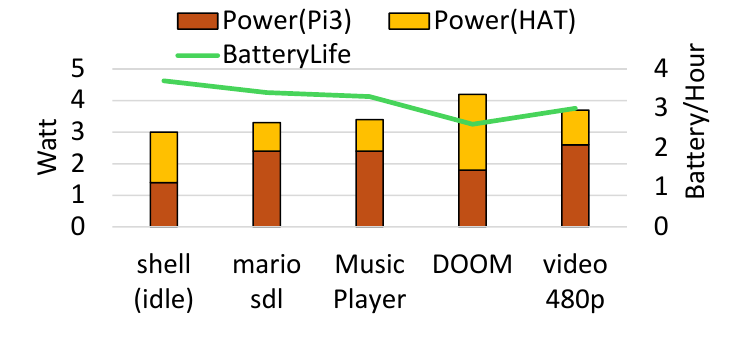}
	\caption{\textbf{Measured device power and estimated battery life}. Power draw broken down to that for Pi3 board as well as for an extension board (HAT) that offers a display, sound amplifier, and power IC. 
	Battery life estimated based on the combined power and one 18650 battery (3000 mAH 3.7V) compatible with the HAT.}
	\label{fig:power}
\end{figure}


\subsection{Power efficiency and battery life}
\label{sec:eval:power}

We evaluated the system's power consumption on real hardware to ensure acceptable battery life for portable demos. 
Measurements were taken on a Raspberry Pi 3 with a GAMEHAT extension board (3.5-inch IPS display), 
both powered by a metered USB power supply. 
The display used its default backlight level.
As shown in~\autoref{fig:power}, the system draws about 3W at shell prompt, yielding an estimated 3.7 hours of battery life. 
Under load (e.g., running \texttt{mario-sdl} or \texttt{DOOM}), consumption rises to $\sim$4W, with battery life dropping to $\sim$2.6 hours. These results confirm the system is power-efficient enough for demonstration use.



\begin{figure}
	\centering
	\includegraphics[width=0.48\textwidth]{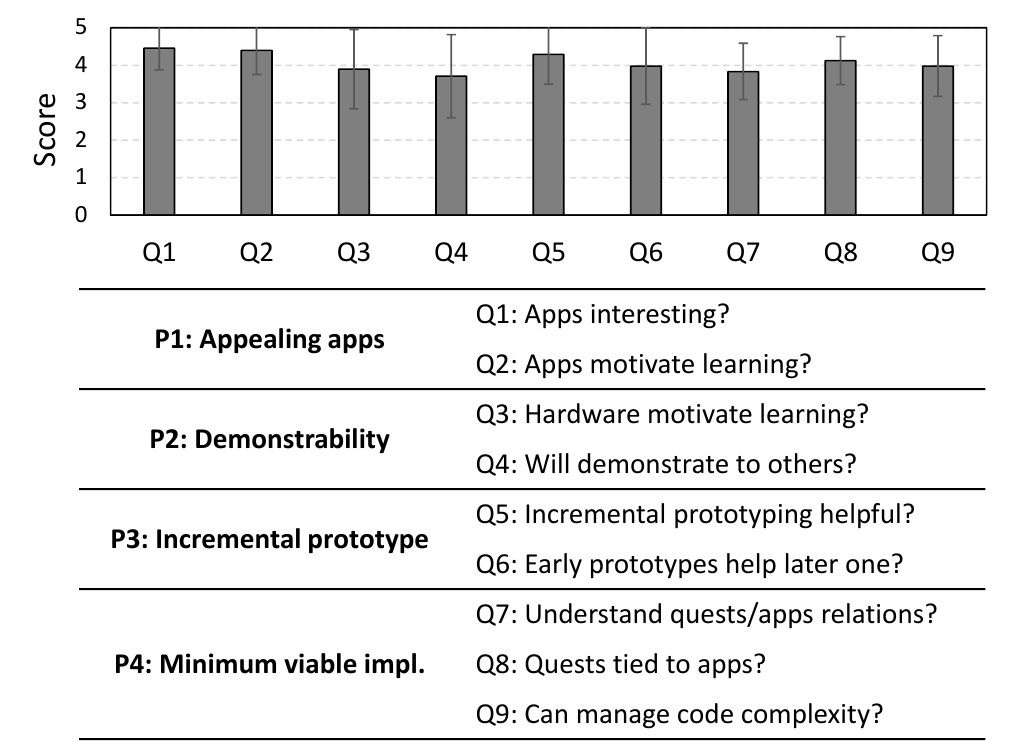}
	\caption{\textbf{A pedagogical survey on \sys{}}, validating our design principles P1--P4 (\S\ref{sec:intro}). 
	Responses on a scale of 1 (strong disagreement) to 5 (strong agreement). N=48.
	}
	\label{fig:survey}
\end{figure}
\subsection{Pedagogical user study}
\label{sec:eval:user}

In Spring 2025, we offered an OS course centered around \sys{} for the first time. Toward the end of the semester, we conducted a survey and received 48 responses out of 59 enrolled students, who were a mix of computer science and engineering majors.

\paragraph{Quantitative results}
\autoref{fig:survey} summarizes students' ratings on how \sys{}'s four principles (P1--P4, \S\ref{sec:intro}) directly supported their learning. (P1) Most students found the interactive apps to be a strong motivator. Many appreciated the opportunity to showcase their apps to others. (P2) Although students could choose between working on real hardware or using emulators, a majority (64\%) opted to experiment with real devices and were motivated. Some students noted challenges in setup and debugging. (P3) Incremental prototyping was effective, with early prototypes serving as clear stepping stones to later ones. (P4) Students reported a clear understanding of the relationship between apps and their dependencies, as well as the connections between different prototypes. Overall, the survey validates the pedagogical value of \sys{}.

\paragraph{Qualitative results}
Students provided open-ended feedback on \sys{} and the course.
They praised the following aspects:
(1) strong application motivations (“Five OS prototypes… culminating in DOOM… kept me motivated”);
(2) incremental complexity across prototypes (“The progressive nature of the labs… created a satisfying learning arc”);
(3) diverse learning paths (“Choose-your-own-adventure format made labs digestible”); and
(4) tight lecture–lab integration (“One without the other would’ve left me confused”).
They also reported the following challenges:
(1) difficulty of developing on real hardware (“RPI3 impractical to develop while traveling…”);
(2) a desire for deeper exploration of certain topics (“Labs went for breadth over depth… wished side quests added depth”);
(3) workload pressure (“Painful at times… worth it, but labs took significant time”); and
(4) a need for more scaffolding in early labs (“Jump from basics to implementation was steep”).
We will address this feedback in future course iterations.

\paragraph{Discussion}
Looking ahead, we plan to continue refining the prototypes and incorporating feedback from students. To make the learning curve more approachable, we aim to enhance the documentation for each prototype, offering clearer background, motivation, and references to support student understanding. To broaden the learning experience, we plan to develop additional applications that integrate diverse kernel features, expand the set of lab branches so students can pursue different paths, and provide a curated list of project options for students to explore in their final work.

\section{Conclusions}

\sys{} showcases that building a modern, usable OS can be both demonstration-ready and educational. 
By tying modern OS features to real apps, we keep the OS construction a motivated journey, with a clear purpose behind system design. 
The result is a small, efficient kernel that supports rich apps and runs on real hardware. 
\sys{} makes OS construction more approachable for a wider audience, without losing depth or realism.

\section*{ACKNOWLEDGMENT}
The authors were supported in part by NSF awards \#2128725, \#1919197, \#2106893, and \#2426353. The authors thank the anonymous reviewers and the shepherd, Eddie Kohler,  for their insightful feedback. 


\bibliographystyle{plain}
\bibliography{bib/abr-short,bib/wrx, bib/ab, bib/misc}

\appendix
\section{Appendix}
We include the task graphs for labs 1–5 below.

\begin{figure*}[htbp]
    \centering

    \begin{subfigure}{0.37\textwidth}
        \centering
        \includegraphics[width=\linewidth]{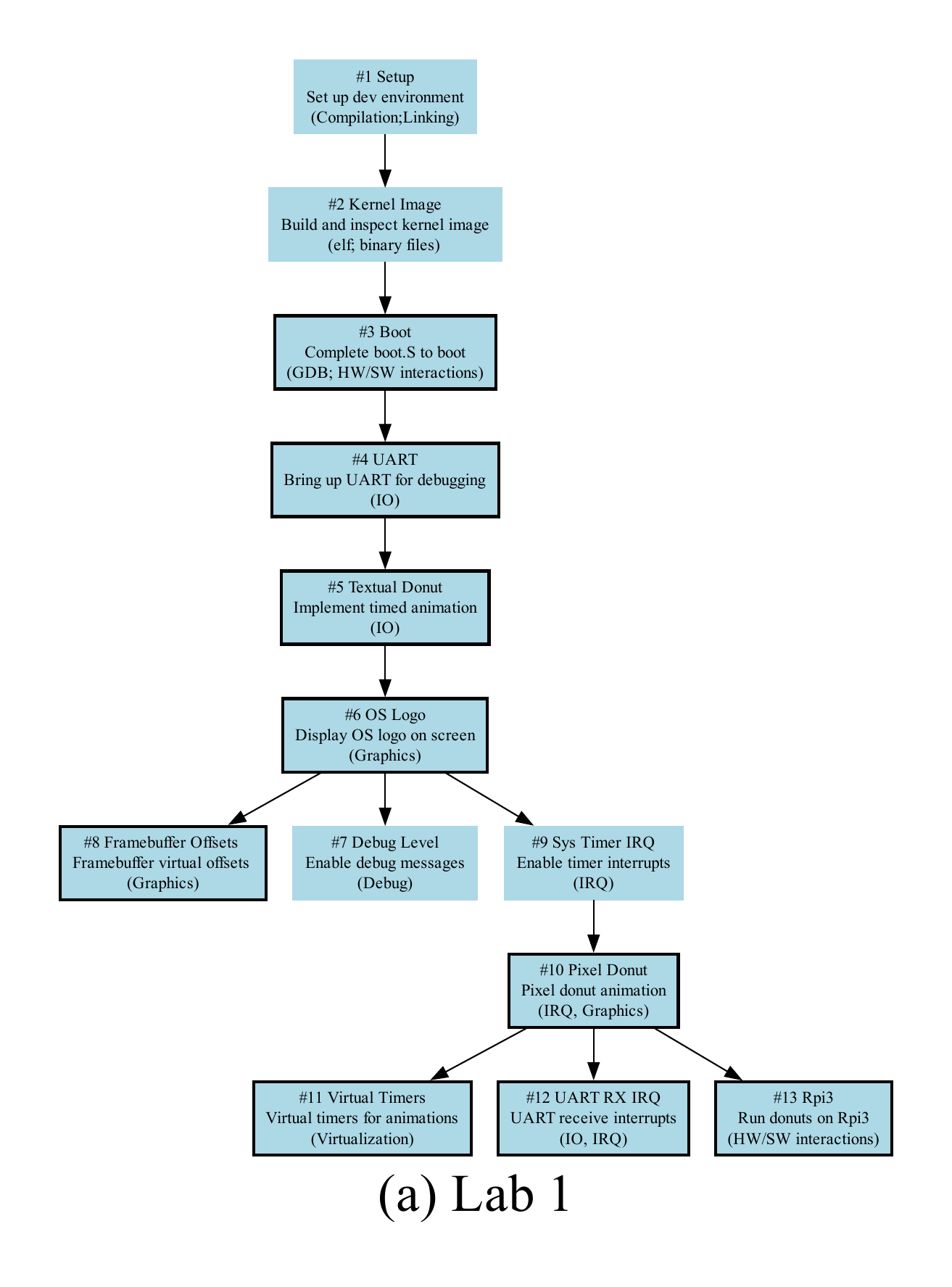}
        \label{fig:lab1}
    \end{subfigure}
    \hfill 
    \begin{subfigure}{0.37\textwidth}
        \centering
        \includegraphics[width=\linewidth]{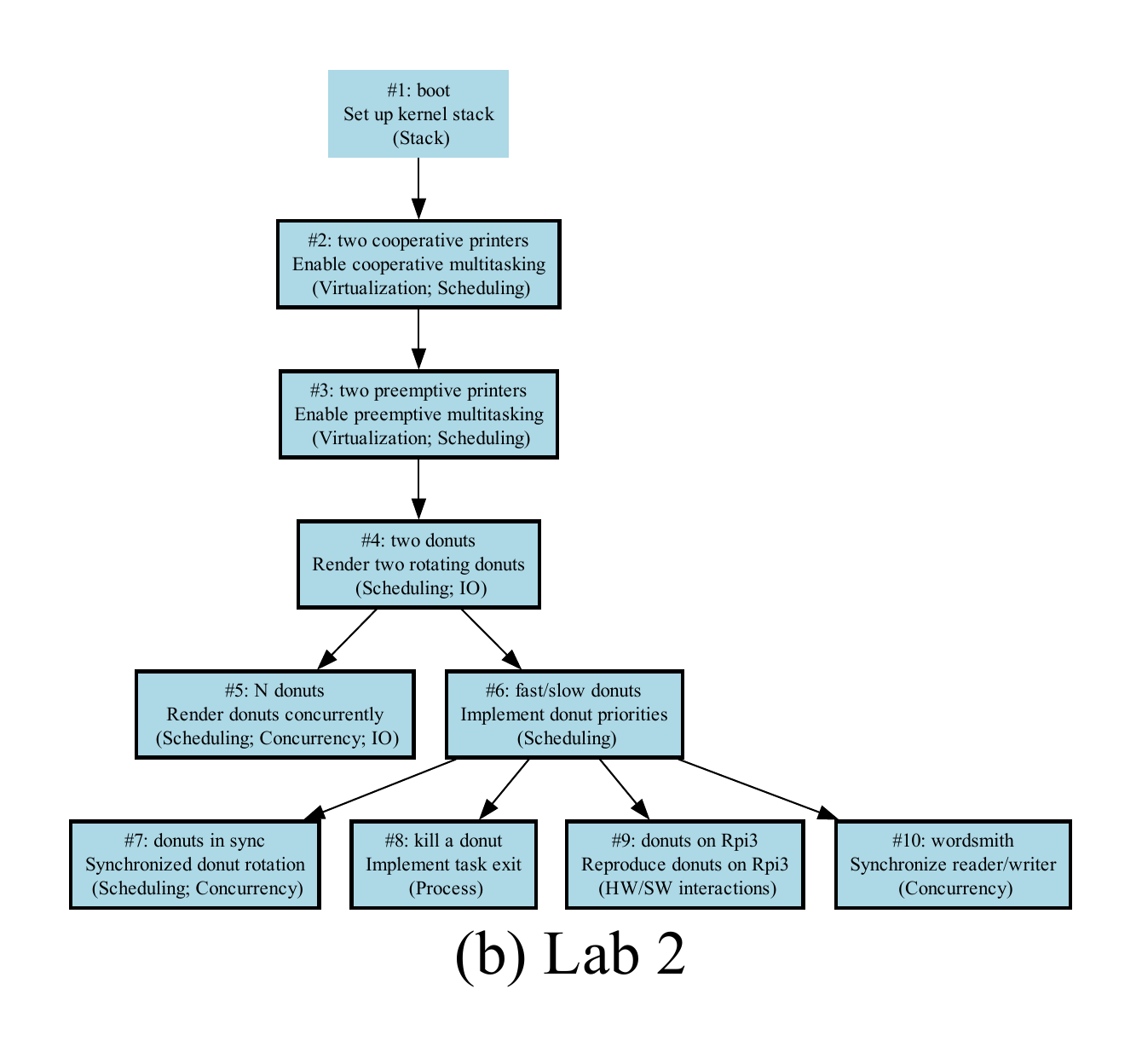}
        \label{fig:lab2}
    \end{subfigure}

    \vspace{1em} 

    \begin{subfigure}{0.37\textwidth}
        \centering
        \includegraphics[width=\linewidth]{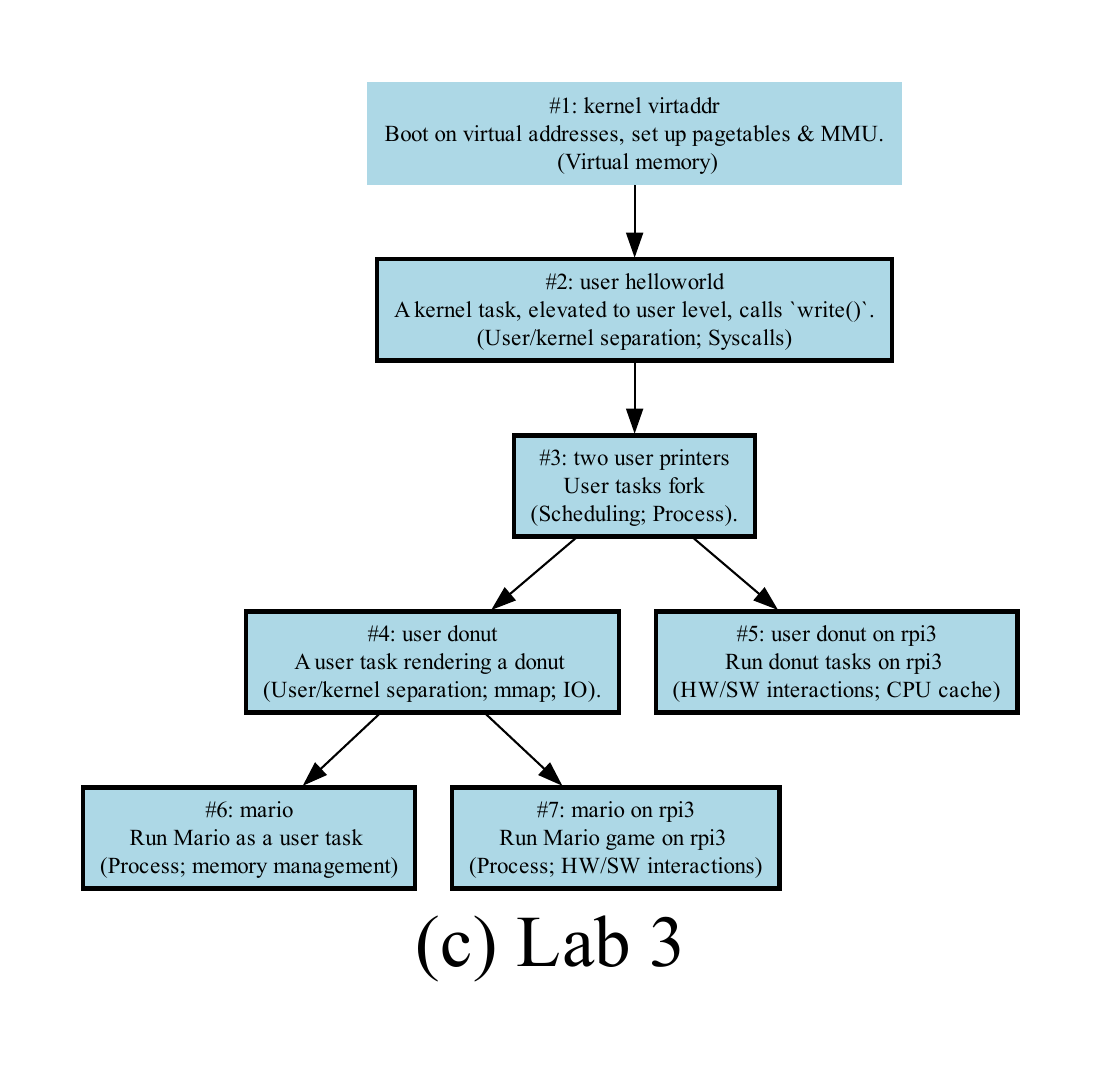}
        \label{fig:lab3}
    \end{subfigure}
    \hfill
    \begin{subfigure}{0.37\textwidth}
        \centering
        \includegraphics[width=\linewidth]{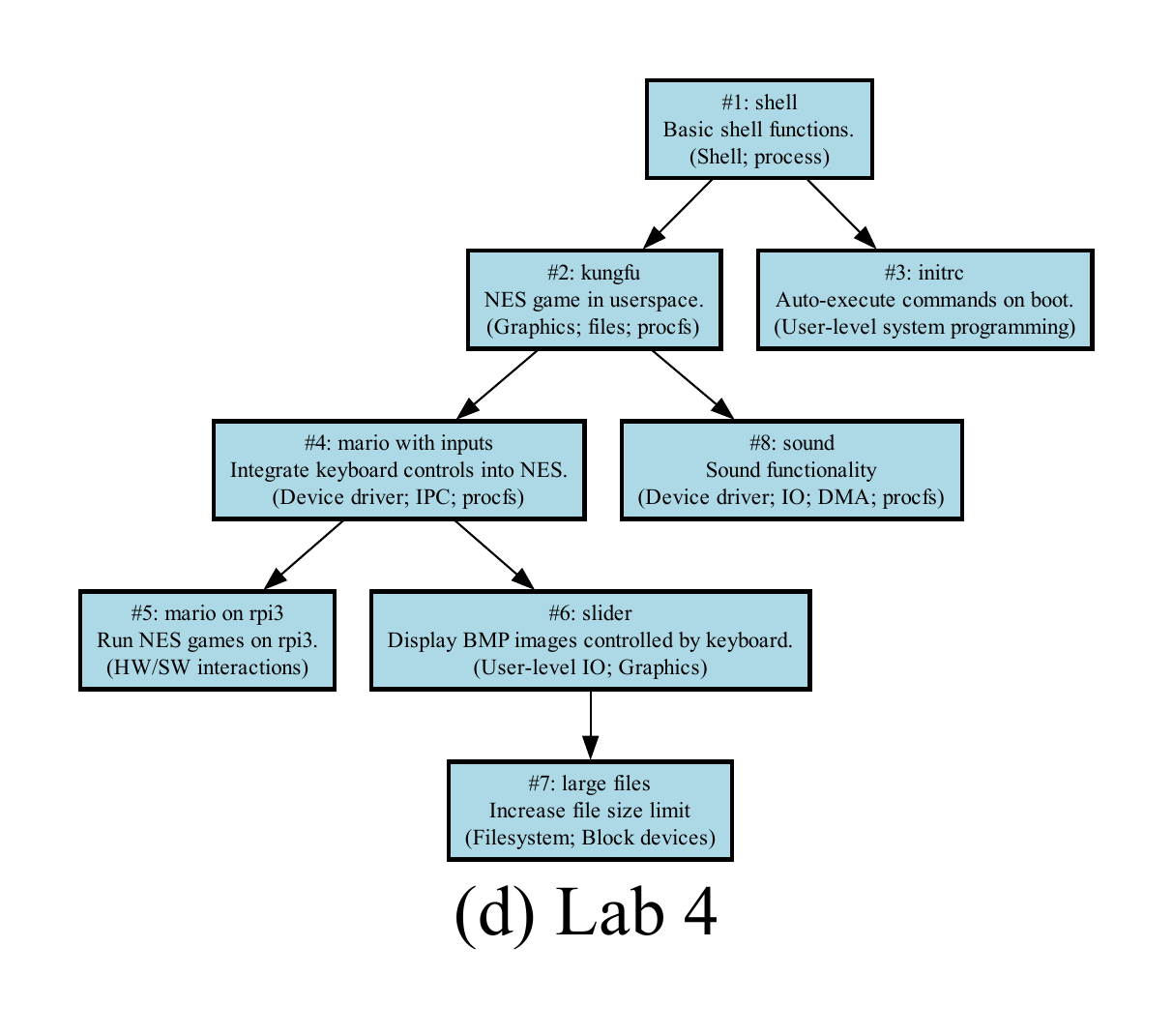}
        \label{fig:lab4}
    \end{subfigure}

    \vspace{1em}

    \begin{subfigure}{0.5\textwidth}
        \centering
        \includegraphics[width=\linewidth]{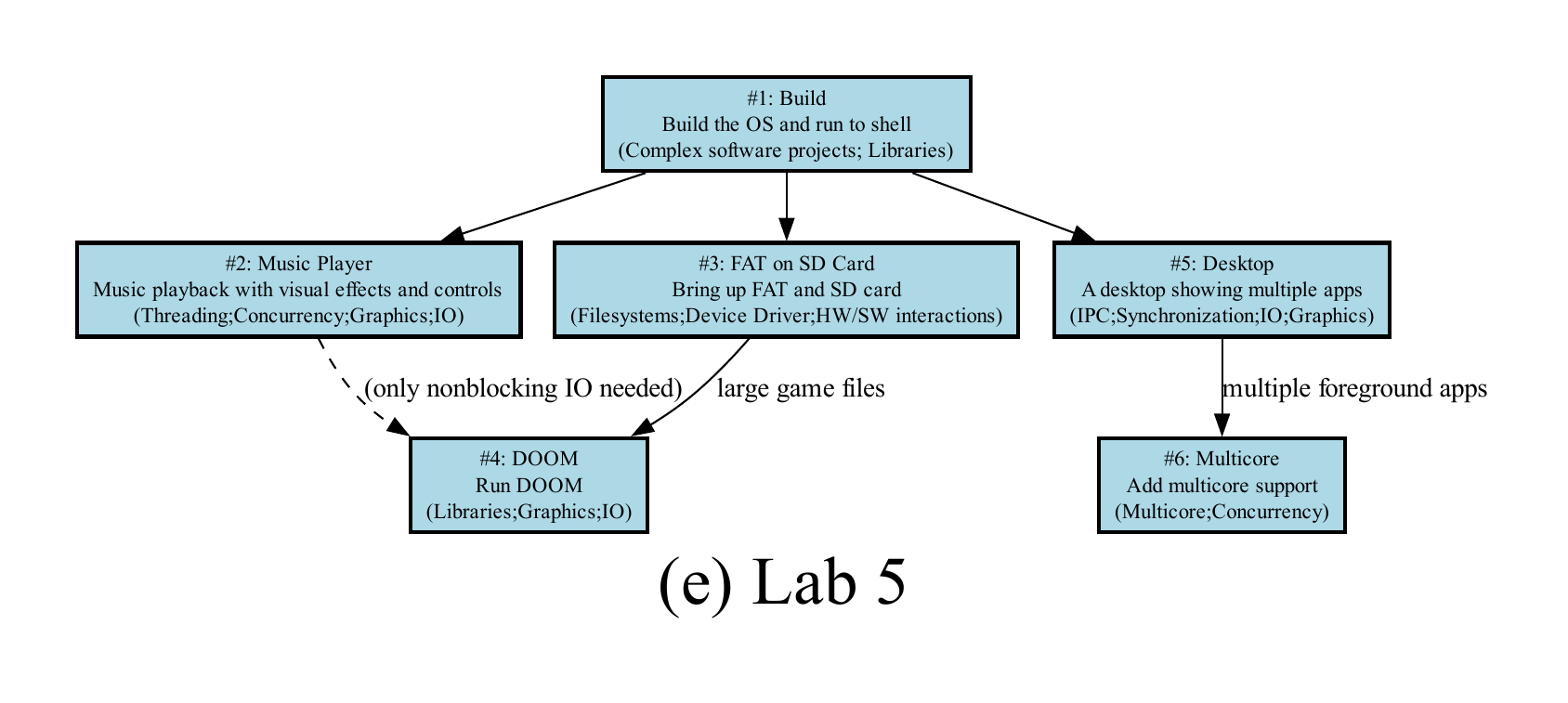}
        \label{fig:lab5}
    \end{subfigure}

    \caption{Task graphs for all labs. Each box represents a task, with the corresponding OS concepts indicated in parentheses. Boxes with borders denote tasks for which students must submit video evidence for assessment.}
    \label{fig:assignments}
\end{figure*} 



\end{document}